\documentclass[preprint,12pt]{elsarticle} 
\usepackage{graphicx}
\usepackage{multirow}
\usepackage{amssymb}
\usepackage{lineno}

\def\MeV{\,{\rm MeV}}
\def\GeV{\,{\rm GeV}}

\def\um{\,\ensuremath{\mu{\rm m}}}
\def\mm{\,{\rm mm}}

\def\m{\,{\rm m}}

\def\AA{\,\hbox{\accent'27A}}				
\let\and=\AA								

\def\mrad{\,{\rm mrad}}

\newcommand{\beq}{\begin{eqnarray}}
\newcommand{\eeq}{\end{eqnarray}}

\newcommand{\slitw}{\ensuremath{w_{\mbox{slit}}}}
\newcommand{\sx}{\ensuremath{\sigma_x}}
\newcommand{\sxp}{\ensuremath{\sigma_{x'}}}
\newcommand{\wscreen}{\ensuremath{w_{\mbox{screen}}}}

\journal{Nuclear Physics A}

\begin{document}

\begin{frontmatter}



\title{
{\Large \bf 
Single Shot Transverse Emittance Measurement of multi-MeV Electron Beams \\ Using a Long Pepper-Pot
}
}
 
\author{
Nicolas Delerue\footnote{Now at LAL, Universit\'e Paris-Sud, CNRS/IN2P3, F-91898 Orsay, France}
}

\ead{delerue@lal.in2p3.fr}


\address{John Adams Institute for Accelerator Science,\\University of Oxford\\OX1 3RH Oxford\\United Kingdom}

\begin{abstract}


We present a pepper-pot design  in which we address the problem of penetration by high energy particle, deriving analytical expressions and performing GEANT4 simulations for the estimation of the error introduced by a long (thick) pepper-pot. We also show that a careful design allows to measure the emittance of electron beam of several hundred MeV and beyond. 

\end{abstract}


\end{frontmatter}


\section{Introduction} 

Laser-driven Wakefield acceleration is a rapidly expanding field which has experienced a  breakthrough in the last few years in the sense that several experiments around the world have achieved beams in the range of hundreds of MeV energy~\cite{Rowlands-Rees:2008,PhysRevLett.103.035002,PhysRevSTAB.13.031301} or more~\cite{leemans2006,1367-2630-9-11-415} over only a few millimetres, achieving gradients as high as 10-100~GeV/m. 

Although such beams could have very interesting applications~\cite{dinoLS,Gruner:2006pw} their shot to shot stability is not yet good enough to allow them to be used in routine operation delivering a  beam for users. In order to study their stability and their characteristics, single shot diagnostics techniques need to be developed. In particular, the measurement of the emittance is essential. To this aim we have re-designed and re-developed the so called ``pepper-pot'' method applied to the high energy electron beam, up to several hundreds MeV. In the design that is presented in this paper, we address the problem caused by the penetration depth of high energy particles, which can prevent the pepper-pot to be used as a reliable diagnostic.

\section{Transverse emittance measurement at low energy} 

\begin{figure}[htbp]
\begin{center}
\includegraphics[height=3cm]{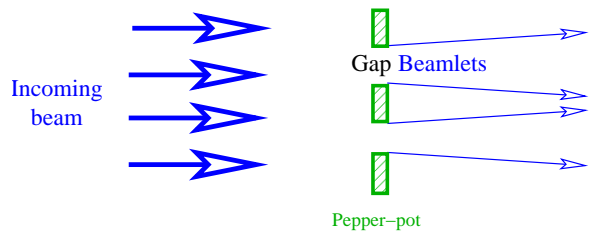}
\caption{
Pepper-pot sampling a beam into several beamlets.
\label{fig:pepperpot} }
\end{center}
\end{figure}

The pepper-pot method is often used at low energy to measure the transverse emittance of particle beams~\cite{Lejeune,HumphriesCPB,zhang1996emittance}. In this method an array of holes in a sheet of material is used to separate a large beam into several beamlets (see figure~\ref{fig:pepperpot}). After a short drift length the width of each beamlet can be measured. This width gives a measure of the beam divergence at the position where the beam was sampled by the pepper-pot. Hence pepper-pots can be used to measure simultaneously the transverse size and the divergence of each beamlet thus providing a direct measurement of the beam transverse emittance.

With a two dimensional array of holes the two components of the transverse emittance can be measured simultaneously. If only one component of the transverse emittance is to be measured this 2D-array of holes can be replaced by a 1D array of slits. For simplicity this paper will focus on a 1D array but the arguments developed here could also be applied to a 2D array.

\section{Penetration of high energy electrons into matter} 
\label{sec:geant}
\begin{figure*}[htbp]
\begin{center}
\includegraphics[height=8cm]{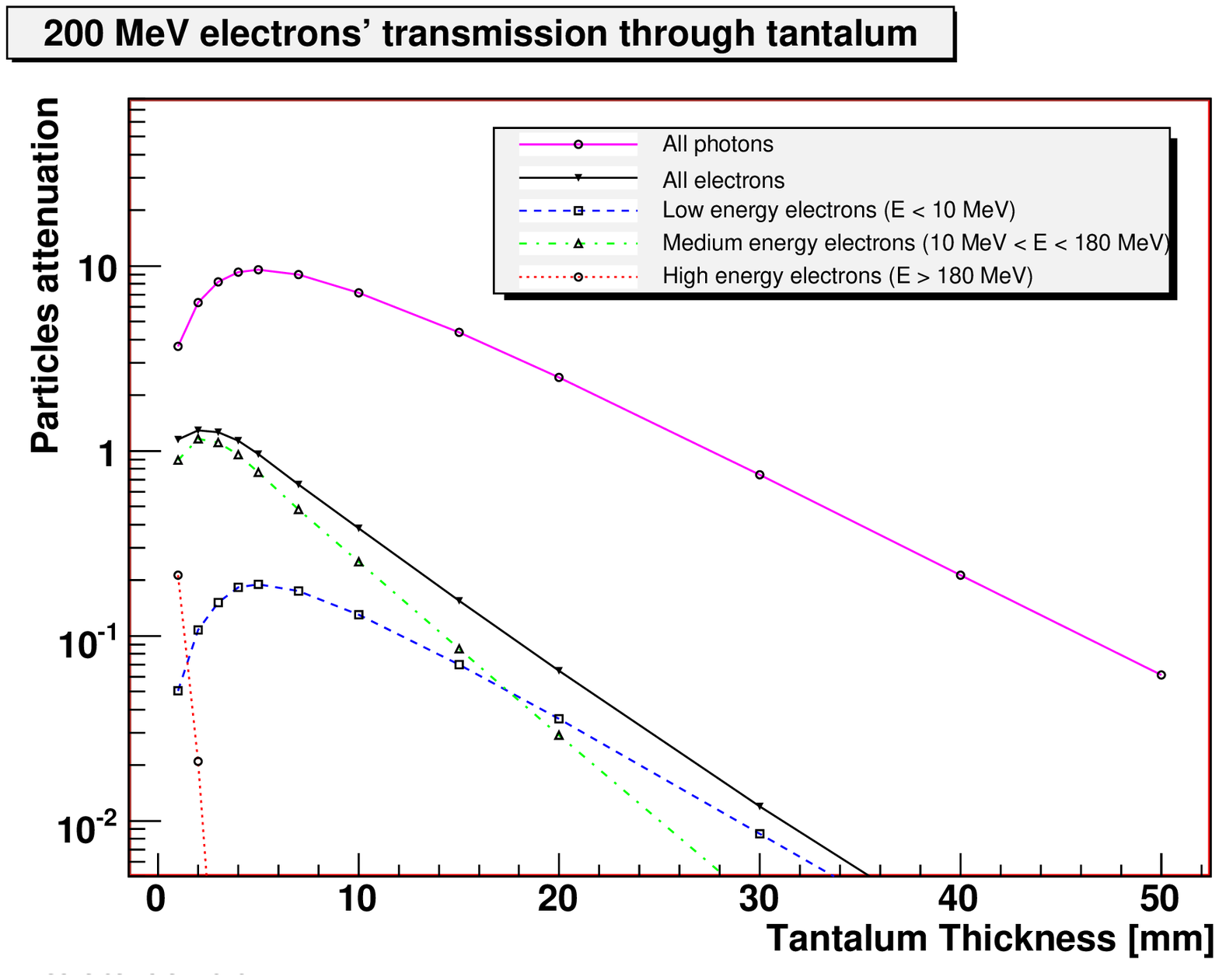}
\includegraphics[height=8cm]{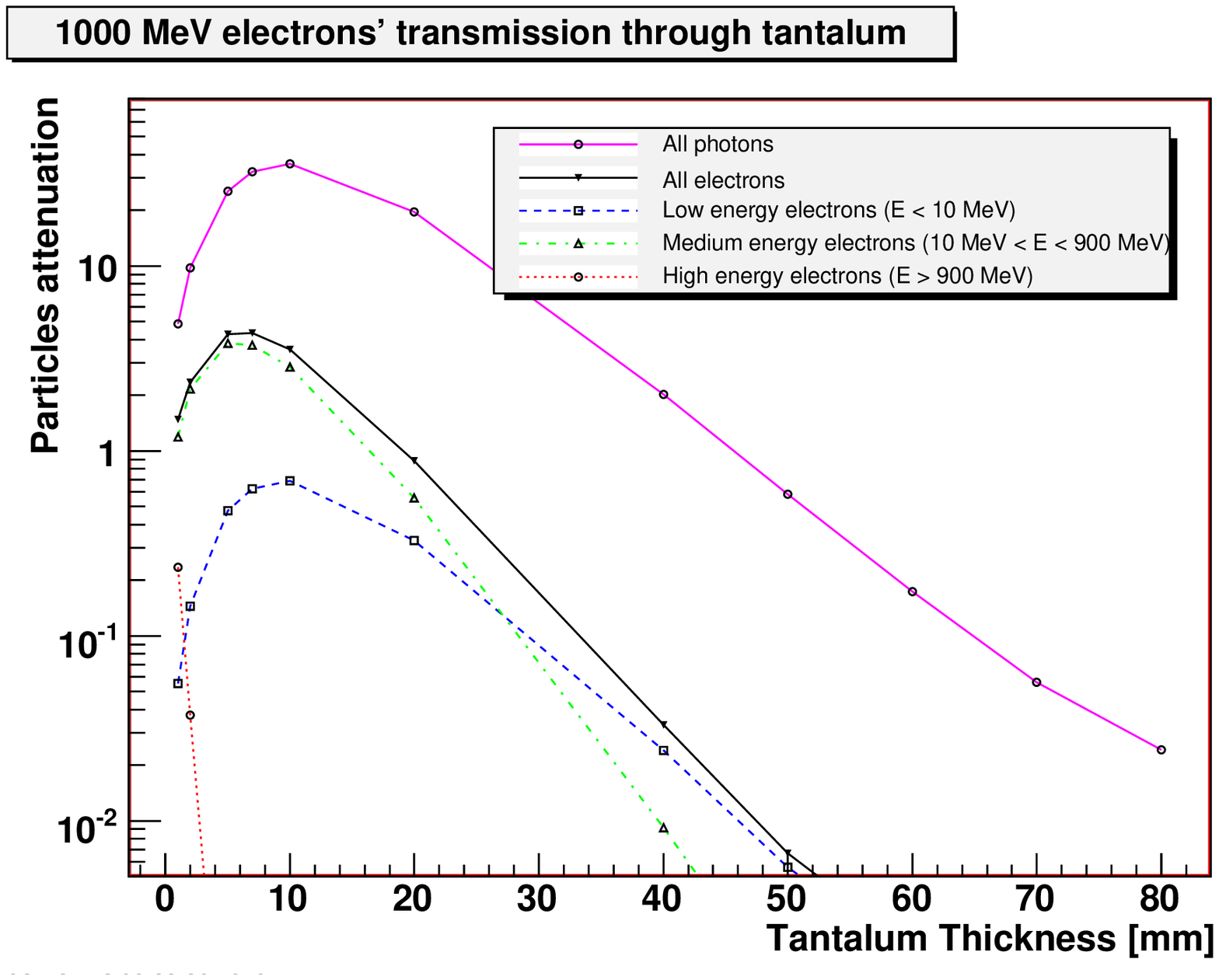}
\caption{
GEANT4~\cite{Agostinelli:2002hh} simulations of the penetration of high energy electrons in a block of Tantalum. The horizontal axis gives the Tantalum thickness and the vertical axis gives the number of electrons that will be seen by a circular detector located 1.5m downstream. The vertical axis is normalised to one initial electron. 
The red lines (round markers) correspond to high energy electrons whereas the blue lines (square markers) and the green lines (up-pointing triangular markers) correspond to lower energy particles emitted by the initial particle. The black lines (down-pointing triangular markers) show the total number of electrons remaining. The magenta lines (round markers)  show the total number of photons at that location.
The plot on the top corresponds to 200\MeV\  electrons and the one on the bottom to 1\GeV\  electrons.
\label{fig:geant} }
\end{center}
\end{figure*}

The main problem of using a Pepper-pot as diagnostic for transverse emittance measurement of high energy particles, arises from the thickness of material required to give sufficient contrast between the beam passing through the hole and the stopped beam.
The thickness of material needed to stop electrons with an energy of several hundreds of MeV is quite large. Although one radiation length is enough to absorb in average all but $1/e$ of the electrons' energy~\cite{PDG}, the GEANT4~\cite{Agostinelli:2002hh} simulations shown in figure~\ref{fig:geant} indicate that several additional radiation lengths are required to stop most of the electronic shower.


\begin{figure*}[htbp]
\begin{center}
\includegraphics[height=4cm]{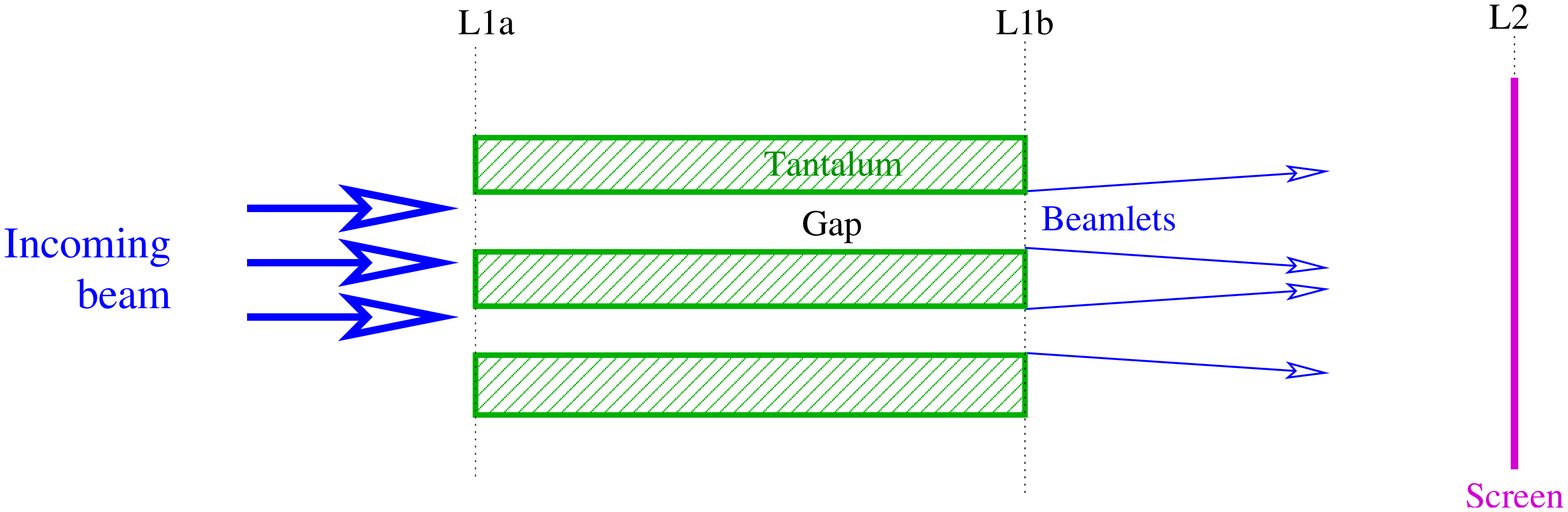} 
\caption{
Long pepper-pot: the slits or holes are longer than with an usual pepper-pot.
\label{fig:extended_pepperpot} }
\end{center}
\end{figure*}

Therefore, a pepper-pot able to measure the transverse emittance of a beam of several hundred MeV electrons must have a significant longitudinal extent (figure~\ref{fig:extended_pepperpot}) and one may question how the acceptance of the pepper-pot is modified by this length.

\begin{figure*}[htbp]
\begin{center}
\includegraphics[height=6cm]{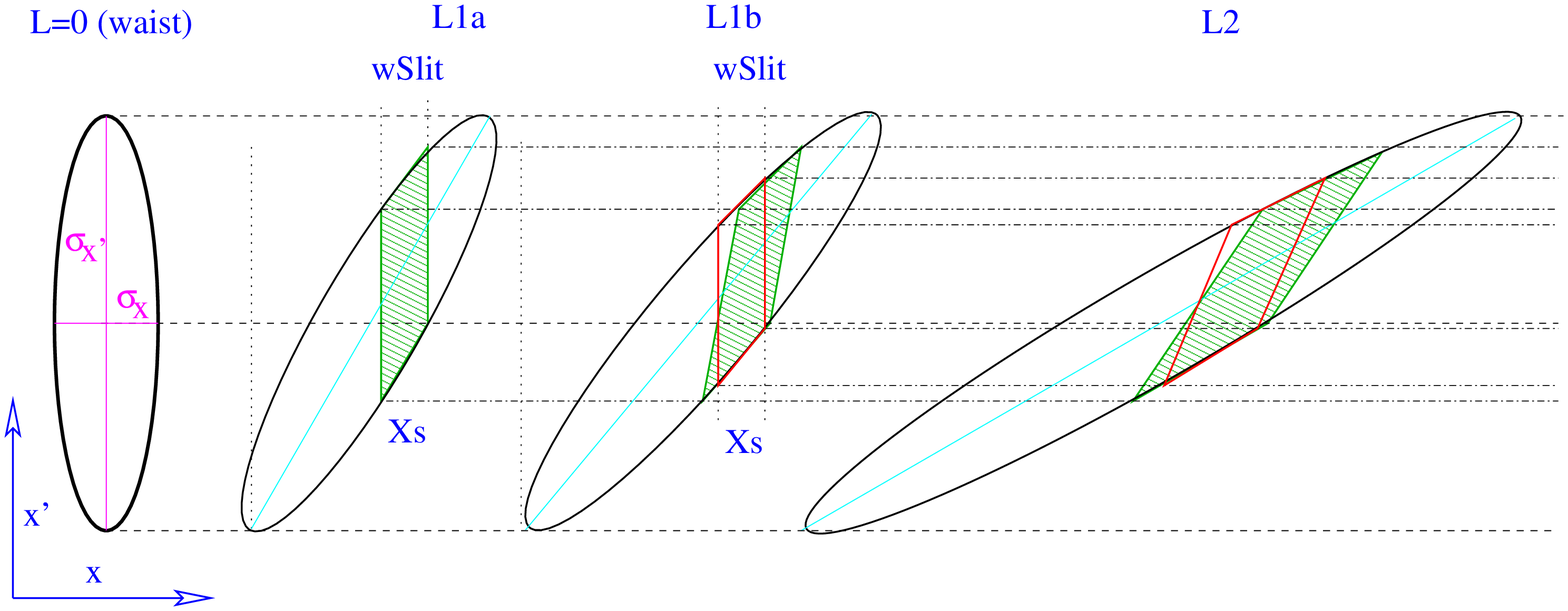} 
\caption{
Long pepper-pot: Evolution of the phase-space as the beam drifts. At the source or at a wait ($L=0$) the phase space of the beam is first represented by an upright ellipse. As the beam drifts the ellipse gets sheared. At $L_{1a}$ the beam enters a slit and thus only the particles in the green area pass the slit. While the beam travels in the slit the ellipse is further sheared, hence at $L_{1b}$, the exit of the slit, the area of phase space within the slit acceptance (red box) is different from that occupied by the particles sampled at the entrance of the slit. Hence the full acceptance of the slit correspond to the overlap between the green and the red areas.
\label{fig:extended_pepperpot_phase_space} }
\end{center}
\end{figure*}

Figure~\ref{fig:extended_pepperpot_phase_space} shows how the phase space of the beam evolves as the beam drifts. At the entrance of a long slit only the particles with a position in a certain range ($ X_s < x_{1a} < X_s + \slitw $) will enter the slit. While the beamlet travels in the slit these particle will propagate and at the exit they will have a different position. However the slit is still limiting particles to the same range ($ X_s < x_{1b} < X_s + \slitw $). To understand how the length of the pepper-pot affects its resolution it is interesting to calculate the ratio between the area sampled by both the entrance and the exit of the slit with the area sampled by either the entrance or the exit of the slit.

To perform this calculation we will first consider the effect on the beam of a narrow slit (section~\ref{sec:narrow_slit}) then the effect of a wide slit (section~\ref{sec:wide_slit}) and in section~\ref{sec:long_slit} the effect of a long slit. 
Finally in section~\ref{sec:parametres} will discuss how the positions of the pepper-pot and of the screen should be chosen.

In this paper ``narrow'' or ``wide'' refer to the dimensions transverse to the direction of propagation of the beam whereas ``short'' or ``long'' refer to the dimension of propagation of the beam.


To make the calculations in this study simpler we will approximate the phase space ellipse  by a parallelogram. This approximation is valid as long as the beam is large with respect to the size of the slits (For the purposes of this paper we ignore the beam energy spread and thus there is no distinction between trace space and phase space). 

At a waist, such as the source of the beam or a focusing device, the ellipse (parallelogram) is upright but as the beam travels in drift space this countour will be sheared.

The definition of variables is as follows:
\begin{itemize}
\item $L_0$ denotes the location of the beam waist. $L_1$, $L_{1a}$, $L_{1b}$ and $L_2$ are other locations at which the beam is observed.
\item \sx\  and \sxp\  are the half-width and half-divergence of the beam at location $L_0$ 
\item $x_i$ and $x'_i$ are the position and the divergence of a given particle at location $L_i$.
\item $A_i$ ($B_i$) is the position in the phase space of the particle with the highest (lowest) divergence at location $L_i$.
\item $X_s$ is the position of the slit, $d_s$ is the divergence of the particles selected by this slit and \slitw is the width of this slit. 
\end{itemize}


\section{Effect of a narrow slit} 
\label{sec:narrow_slit}

\begin{figure*}[htbp]
\begin{center}
\includegraphics[width=10cm]{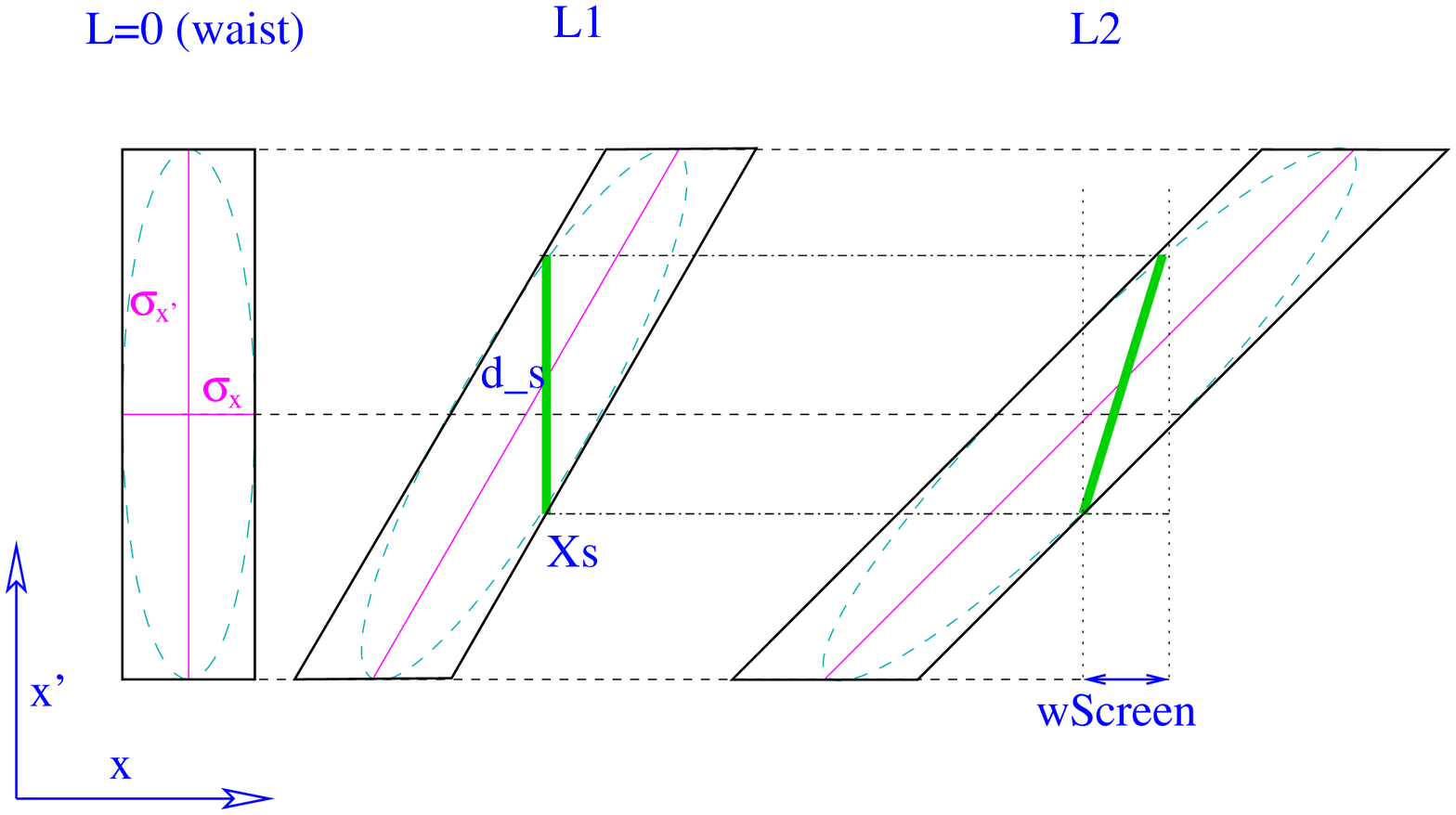} 
\includegraphics[width=10cm]{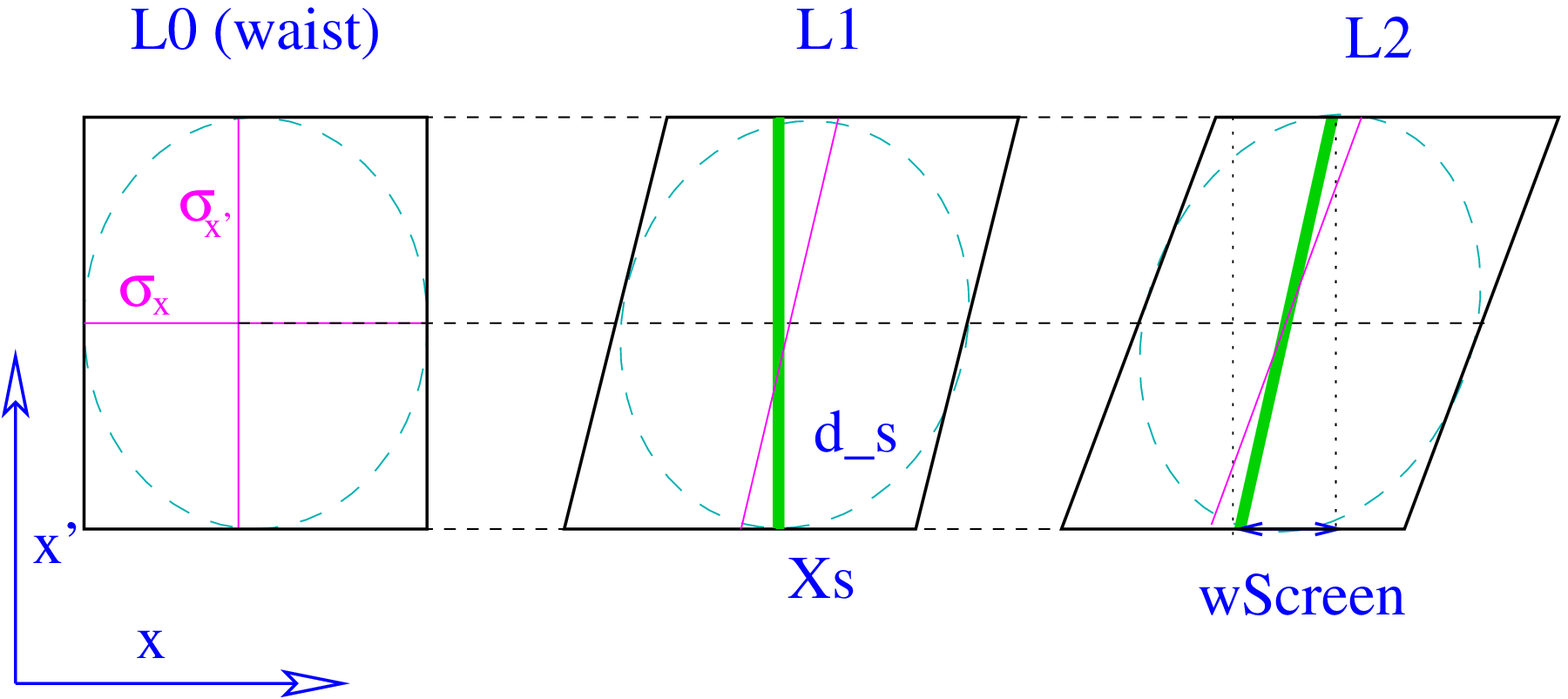} 
\includegraphics[width=10cm]{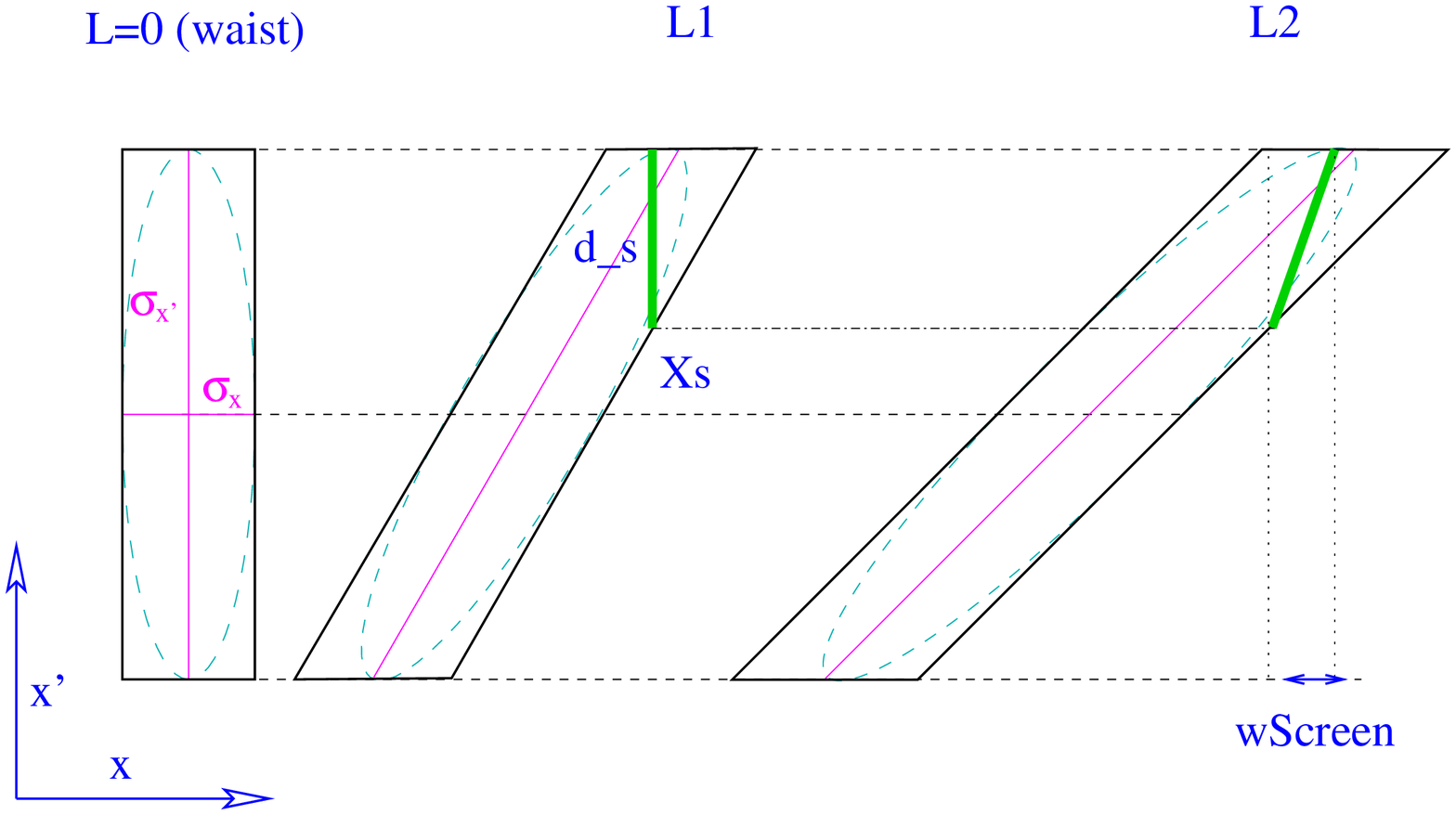} 

\caption{
Phase space of a beam sampled by a narrow slit as the beam propagates in a drift space. The emittance ellipse is drawn in dashed blue and its approximation by a rectangle is drawn in black. The top figure shows the case where the size of the beam at the slit is dominated by its divergence. The middle figure shows the case where the size of the beam at the slit is dominated by its original size. The bottom figure shows the intermediate case where one end of the pattern is dominated by the divergence and the other end is dominated by the original size.
\label{fig:ellipses_narrow} }
\end{center}
\end{figure*}

To understand what is the effect of each hole or slit, we will consider first 
the effect of one slit imaged on a screen.
Each narrow slit samples the beam by selecting only the particles at a given position ($X_s$). At location $L_1$ 
the slit will select all particles with $x=X_s$ (see figure~\ref{fig:ellipses_narrow}). Let's investigate the shape of the phase space selected by this slit when imaged on a screen at location $L_2$.

To be at $x_1=X_s$ at $L_1$ the position ($x_0$) and divergence ($x'_0$) of the particle at the origin ($L_0$) must satisfy $X_s = x_0 + L_1 x_0'$. 
Thus at $L_2$ all particles passing the slit will satisfy the relation: $x_2 = X_s + (L_2 - L_1) x_2'$ and given that $x'_2 = x_0'$ we have the relation 
\beq
x_2 & = & X_s + (L_2 - L_1) x'_0
\eeq

This defines a line in the phase space as shown in figure~\ref{fig:ellipses_narrow}.

In order to estimate the width $\wscreen$ of this line as measured on a screen at $L_2$ we consider two different cases\footnote{For simplicity we assume $X_s > 0$. The case $X_s<0$ is symmetric to the one described here.}: 
\begin{itemize}
\item A case where the beam size at the slit is dominated by the contribution ($\sxp L_1$) 
of its divergence and therefore $\sxp > \frac{\sx + X_s}{L_1}$ or $|X_s| < ( \sxp L_1 ) - \sx$ (this case is shown on the upper part of figure~\ref{fig:ellipses_narrow}).
\item A case where the beam size at the slit is dominated by the contribution ($\sx + X_s$) and therefore $\sxp < \frac{\sx + X_s}{L_1}$ or $|X_s| < \sxp L_1 + \sx $ (this is shown on the middle part of figure~\ref{fig:ellipses_narrow}).
\item The intermediate case $\frac{X_s - \sx}{L_1} < \sxp < \frac{\sx + X_s}{L_1}$ or $ | X_s \pm \sxp L_1 | < \sx $ can be derived from these two cases. It corresponds to patterns for which one end will be dominated by by the contribution ($\sxp L_1$) of the divergence whereas the other end will be dominated by the contribution $\sx + X_s$ (this is shown on the lower part of figure~\ref{fig:ellipses_narrow}).
\end{itemize}

For all three cases we can write:
\beq
- \sx & < x_0 < & \sx \nonumber \\
X_s & = & x_0 + L1 x'_0 \nonumber \\
max\left(\frac{X_s - \sx}{L_1}, -\sxp \right) & < x'_0 < & min\left( \frac{X_s + \sx}{L_1} , \sxp \right) \label{eq:divergenceAtSource} 
\eeq

If the beam divergence is large enough so that $\sxp > \frac{\sx + X_s}{L_1}$, we can write:
\beq
(L_2 - L_1) \frac{X_s - \sx}{L_1}  & < x_2 - X_s < & (L_2 - L_1)  \frac{X_s + \sx}{L_1} \label{eq:maxParticlesPosL2_xs} 
\eeq

The width $\wscreen$ is the total width of the screen covered by particles, hence:

\beq
& \wscreen = &  \frac{2\sx (L_2 - L_1)}{L_1} \nonumber \\
& 2\sx = & \frac{\wscreen L_1}{L_2 - L_1} \label{eq:2w_ws}
\eeq
and the divergence $d_s$ of the beamlet sampled by the slit is
\beq
& d_s = &  \frac{\wscreen}{L_2 - L_1} = \frac{2\sx}{L_1} \label{eq:slitDivergence}
\eeq

else if $\sxp < \frac{X_s - \sx}{L_1}$
\beq
 X_s - L_1 \sxp & < x_0 < & X_s + L_1 \sxp \\
 - (L_2 - L_1) \sxp & < x_2 - X_s < &  (L_2 - L_1) \sxp  \nonumber \\
& \wscreen = & 2\sxp (L_2 - L_1) \\
 2\sxp = & d_s & = \frac{\wscreen}{L_2 - L_1} \label{eq:2d_ws}
\eeq

and if $\frac{X_s - \sx}{L_1} < \sxp < \frac{\sx + X_s}{L_1}$, using ($X_s > 0$) we can write:
\beq
\frac{X_s - \sx}{L_1} & < x'_0 < & \sxp \\
& \wscreen = & (L_2 - L_1) \left( \frac{X_s - \sx}{L_1} - \sxp \right) 
\eeq

It is interesting to note that if the size at the slit is dominatd by the divergence of the beam
 ($\sxp > \frac{\sx + X_s}{L_1}$) the divergence $\sxp$ of the beam has no effect on \wscreen\  and inversely, when the initial size of the beam dominates the size at the slit ($\sxp < \frac{X_s - \sx}{L_1}$) it is the initial size $\sx$ of the beam that has no effect on \wscreen.  The beam produced by laser-driven plasma accelerators typically features a large divergence and a small size and will therefore match the first case whereas conventional gun are more likely to be in the later case.

Hence depending on the regime in which the tests are made, measuring \wscreen\  can either provide information on the initial width (equation \ref{eq:2w_ws}) or on the initial divergence (equation \ref{eq:2d_ws}) of the beam at the source. By repeating this measurement with several slits sampling the emittance ellipse at different positions it is possible to reconstruct the full emittance ellipse of the beam at $L_1$.

\section{Finite width slit} 
\label{sec:wide_slit}

Real slits are not infinitely narrow but have a finite width (\slitw). In this section we will consider how the result of the previous section needs to be modified to take into account this finite width.

\begin{figure}[htbp]
\begin{center}
\includegraphics[height=6cm]{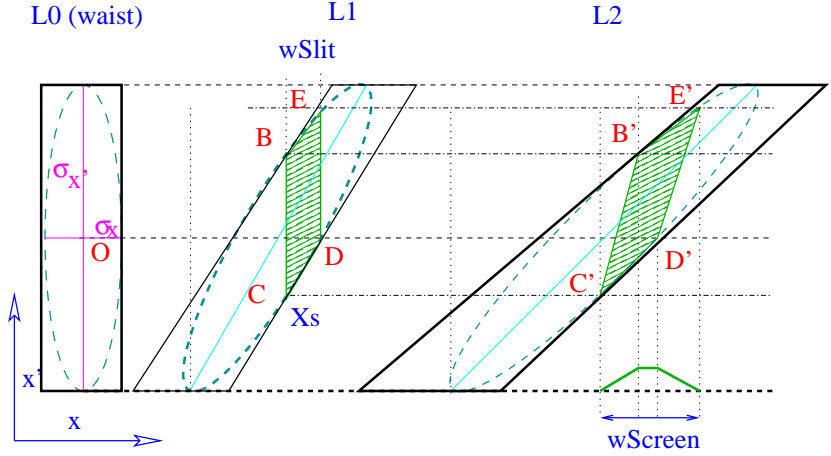} 
\caption{
Emittance ellipse sampled by a slit of finite width (\slitw) as the beam propagates in a drift space.
\label{fig:ellipses_wide} }
\end{center}
\end{figure}

Such a slit will select particles in the range $X_s < x < X_s + \slitw$ (see figure~\ref{fig:ellipses_wide}) and thus the line observed at $L_2$ will be wider than what was found previously.

To satisfy  the condition $ X_s < x < X_s + \slitw  $ at $L_1$ the position ($x_0$) and divergence ($x'=x'_0$) of the particle at the origin ($L=0$) must satisfy $ X_s +\slitw > x_0 + L_1 x_0' > X_s $.
Thus at $L_2$ we have the relation 
\beq
X_s + (L_2 - L_1) x_0' & < x_2 < & X_s + \slitw + (L_2 - L_1) x_0'
\eeq

Once again we must consider two cases.

If $\sxp > \frac{\sx + X_s+\slitw}{L_1}$, to be selected by the slit the particles must satisfy:
\beq
 -\sx & < x_0 < & \sx \nonumber \\
\frac{X_s - \sx}{L_1} & < x_0' < & \frac{X_s + \slitw + \sx}{L_1} \label{eq:divergenceAtSourceWide}
\eeq
hence 
\beq
(L_2 - L_1) \frac{X_s - \sx}{L_1}  & < x_2 - X_s < & (L_2 - L_1)  \frac{X_s + \slitw + \sx}{L_1} + \slitw \label{eq:maxParticlesPosL2_xs_ws} \\
& \wscreen = & 2\sx \frac{L_2 - L_1}{L_1} + \slitw \frac{L_2}{L_1} \\
& 2\sx = & \frac{\wscreen L_1 - \slitw L_2}{L_2 - L_1} \label{2w_wf}  \label{pepperpotRes}
\eeq

Else if $\sxp < \frac{X_s - \sx}{L_1}$, the particles must satisfy:
\beq
 X_s - L_1 \sxp & < x_0 < & X_s + \wscreen + L_1 \sxp
\eeq
hence
\beq
 - (L_2 - L_1) \sxp & < x_2 - X_s < &  (L_2 - L_1) \sxp + \slitw \nonumber \\
& \wscreen = & 2\sxp (L_2 - L_1) + \slitw \\
& 2 \sxp = & \frac{\wscreen - \slitw}{L_2 - L_1} \label{2d_wf}
\eeq

Equations~\ref{2w_wf} and~\ref{2d_wf} show that when the width of the slit is non negligible an extra term must be added to equations~\ref{eq:2w_ws} and~\ref{eq:2d_ws}.

It is also possible to measure the acceptance in the phase space of the slit. At $L_1$, the phase space area $\mathcal{A}_{p}$ can be approximated by a parallelogram. The width of this parallelogram is $\slitw$ and its height is the beam divergence sampled by the slit at $X_s$, it is given by equation~\ref{eq:slitDivergence} (BC in figure~\ref{fig:ellipses_wide}). Hence 
\beq 
\mathcal{A}_p & = & \slitw \times d_s = \frac{2\sx \slitw}{L_1} \label{eq:phase_space_area}
\eeq

By considering the shearing of $B'C'$ and $B'E'$ it can be verified that at $L_2$ this parallelogram has the same area than at $L_1$.

The profile on the screen of each beamlet can also be estimated by looking at the density of particle landing at a given $x$ position on the screen.
The density of the beam is conserved when the beam propagates in a drift space. The parallelogram sampled at $L_1$ has an area $\slitw \times BC = \slitw \times \frac{2 \sx}{L_1} $ (where BC is defined on figure~\ref{fig:ellipses_wide} and assuming the conditions of equation~\ref{eq:maxParticlesPosL2_xs}). Similarly at $L_2$ this area is $\slitw \frac{L_2}{L_1} \times \frac{2 \sx}{L_2}$ which is the same than at $L_1$. However by looking at figure~\ref{fig:ellipses_wide} we see that this is not uniformly distributed. 
Between $x_{B'}$ and $x_{D'}$ the particle density ($pd$) per unit length is $pd(x)=\frac{2 \sx}{L_2}$ (this interval is non-zero only when $x_{B'}  < x_{D'}$):

\beq
pd(x)_{x \in[x_{B'} x_{D'}]} & = & \frac{2 \sx}{L_2} \label{eq:thinPPprofile1}
\eeq

Between $x_{C'}$ and $x_{B'}$ the particle density is raising linearly from $0$ to $\frac{2 \sx}{L_2}$ hence

\beq
pd(x)_{x \in[x_{C'} x_{B'}]} &  & = \label{eq:raisingProfile} \label{eq:thinPPprofile2} \\ & & \frac{1}{L_2} \left[ \left(x - (X_s + \frac{X_s - \sx }{L_{1}}(L_2 - L_1)) \right) \frac{L_{1}}{L_2 - L_1} \right]  \nonumber
\eeq 

From $x_{D'}$ to $x_{E'}$ the particle density is decreasing linearly from $\frac{2 \sx}{L_2}$ to $0$ hence

\beq
pd(x)_{x \in[x_{D'} x_{E'}]} &  & = \label{eq:thinPPprofile3}  \\ & &  \frac{1}{L_2} \left[ \left((X_s + \slitw \frac{X_s + \sx }{L_{1}}(L_2 - L_1)) -x \right) \frac{L_{1}}{L_2 - L_1} \right]  \nonumber
\eeq 

This profile is drawn on figure~\ref{fig:ellipses_wide} under the phase space for $L_2$ and on figure~\ref{fig:beamProfile}.

\section{Long slit with a finite width} 
\label{sec:long_slit}

As discussed in section~\ref{sec:geant}, to be usable at high energies, a pepper-pot with thick slits must be used. With such thick slits the assumption that the beam is being sampled only at $L_1$ is no longer valid. The beam is now being sampled continuously between position $L_{1a}$ and $L_{1b}$.  This will have an effect on the acceptance of the pepper-pot. We consider how this effect can be minimised.

Given that the beams we are interested in usually have a large divergence and a very small source size~\cite{leemans2006} we will consider only the case $\sxp > \frac{\sx + X_s +\slitw}{L_1a}$ in the rest of this paper. This is different from the conditions under which some particle guns are operated.

\subsection{Two consecutive thin pepper-pots}

\begin{figure}[htbp]
\begin{center}
\includegraphics[height=6cm]{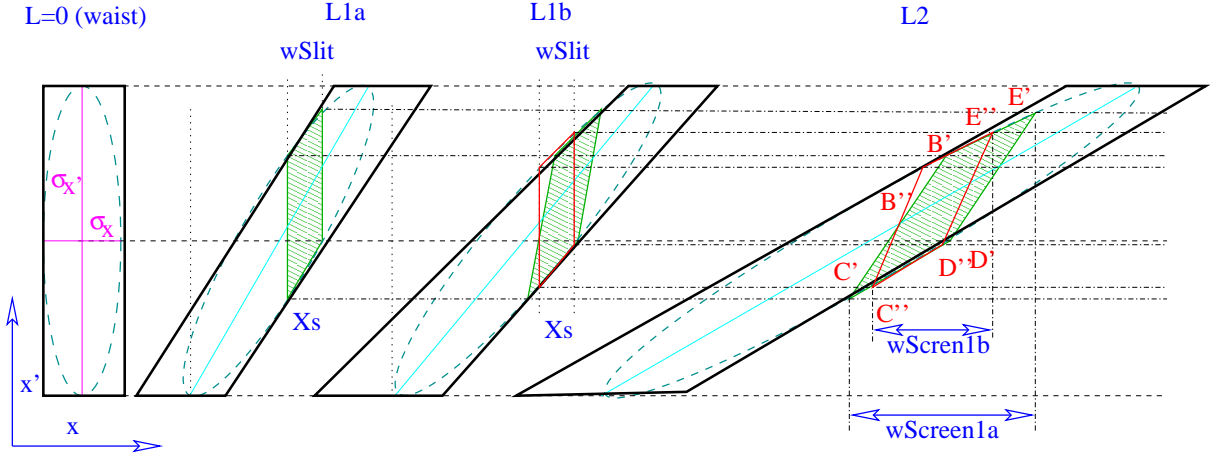} 
\caption{
Emittance ellipses sampled by two slits of finite width ($\slitw$) positioned at $L_{1a}$ and $L_{1b}$.
\label{fig:ellipses_two_PP} }
\end{center}
\end{figure}

We consider two pepper-pots located at $L_{1a}$ and $L_{1b}$ respectively. The first pepper-pot will sample the beam and select particles in the range $X_s < x < X_s + \slitw$ at $L_{1a}$ (see figure~\ref{fig:ellipses_two_PP}).

This sample will then drift toward the second pepper-pot. Some of the particles that were in the range $X_s < x < X_s + \slitw$ at $L_{1a}$ may be outside that range at $L_{1b}$ and other particles may have entered it. For example a particle that passes the first slit at $x_{1a} = X_s + \slitw$ with a positive divergence ($\sxp=\sxp^{+}$) will reach the second slit at $x_{1b} = X_s + \slitw + \sxp^{+} (L_{1b}- L_{1a}) >   X_s + \slitw $ and will thus not pass the second slit. Similarly a particle with a positive divergence ($\sxp=\sxp^{+}$) that should reach the second slit at $x_{1b} = X_s$ would need to reach the first slit at $x_{1a} = X_s -  \sxp^{+} (L_{1b}- L_{1a}) < X_s$ and thus will be blocked by the first slit.

The fraction of the phase space that clears the first slit but is blocked by the second slit can be calculated as follows.

Using the calculation of $\mathcal{A}_{p}$ at equation~\ref{eq:phase_space_area}, we first note that the acceptance in the phase space of the the first slit ($\mathcal{A}_{a}$) is different form that of the second slit ($\mathcal{A}_{b}$):
\beq
\mathcal{A}_{a} & = & \frac{2\sx \slitw}{L_{1a}} \label{eq:areaA} \\
\mathcal{A}_{b} & = & \frac{2\sx \slitw}{L_{1b}}  \label{eq:areaB} \\
\frac{\mathcal{A}_{a}}{\mathcal{A}_{b}} & = & \frac{L_{1b}}{L_{1a}}
\eeq

\begin{figure}[htbp]
\begin{center}
\includegraphics[height=6cm]{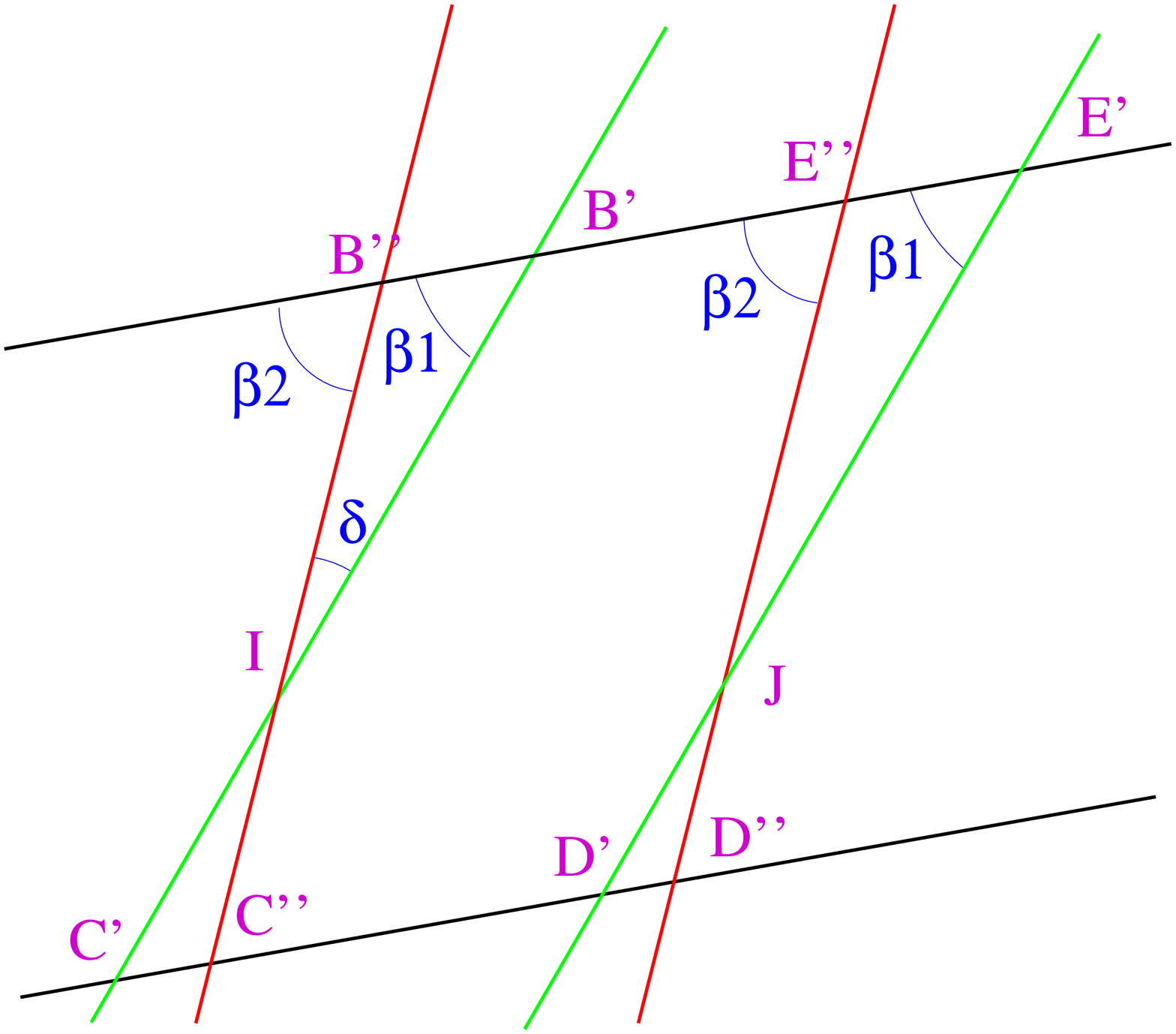} 
\caption{
Enlargement of the area of figure~\ref{fig:ellipses_two_PP} at $L_2$ where the two slits overlap.
\label{fig:parallelogram} }
\end{center}
\end{figure}

Figure~\ref{fig:parallelogram} is an enlargement of the part of figure~\ref{fig:ellipses_two_PP} where the areas of beam sampled by both slits overlap at $L_2$.

At $L_2$ we can estimate the area ($\mathcal{A}_o$) of phase space where the beam sampled by the two pepper-pots overlap. We can compare it with the total area of phase space swept by either slit ($\mathcal{A}_s$).
We find (see~\ref{sec:deriveAreaRation}):

\beq
\frac{\mathcal{A}_{s} - \mathcal{A}_{o}}{\mathcal{A}_{o}} & = & \frac{ \Delta L_{ba} \left[ (X_s - \sx)^2 + (X_s + \slitw + \sx)^2 + (X_s - \sx)^2 + (X_s + \slitw + \sx )^2 \right] }{  2\sx \slitw L_{1b} +  \Delta L_{ba}  \left[ (X_s + \sx)^2 + (X_s + \slitw - \sx)^2 \right]  }   \nonumber \\
\label{eq:areaRatio} 
\eeq

If we consider a central slit ($X_s \simeq 0$) and a correctly aligned pepper-pot ($x_{min} \simeq - x_{max} \simeq \sx$) this equation becomes:
\beq
\frac{\mathcal{A}_{s} - \mathcal{A}_{o}}{\mathcal{A}_{o}} & \simeq & \frac{ 2 (\slitw^2 + 2 \sx^2) \Delta L_{ba} }{2\sx \slitw L_{1b} +   \Delta L_{ba}  \left[ \sx^2 + (\slitw - \sx)^2 \right]}
\eeq

As expected, this equation does not depend on $L_2$: once the phase space has been sampled (after $L_{1b}$) the area of overlap and the area swept will remain constant, regardless of the distance at which the screen is located.

In most applications the long pepper-pot is far from the source and the slits are narrow (that is $ L_{1b} \simeq L_1 >> \Delta L_{ba} >> \slitw$), hence:

\beq
2\sx \slitw L_{1b} & >> &  \Delta L_{ba}  \left[ \sx^2 + (\slitw - \sx)^2 \right] \nonumber \\
\frac{\mathcal{A}_{s} - \mathcal{A}_{o}}{\mathcal{A}_{o}} & \simeq & \frac{ (\slitw^2 + 2 \sx^2) \Delta L_{ba} }{\sx \slitw L_{1}} \label{eq:areaRatioSmpl}
\eeq

Equation~\ref{eq:areaRatioSmpl} implies that when the size of the slit is comparable to that of the beam,  the effect of the pepper-pot separation on the beam phase-space  is only important when they are positioned close to the beam waist. Once the pepper-pots are located at a reasonable distance from the waist the emittance measured is likely to be dominated by other experimental errors.

Here also it is possible to calculate the profile on the screen of each beamlet by looking at the density of particle landing at a given $x$ position on the screen.
If there was only the second slit the particle density would raising linearly between $x_{C''}$ and $x_{B''}$ from $0$ to $\frac{2 \sx}{L_2}$ with an expression comparable to that of equation~\ref{eq:raisingProfile}. However after $x_I$ the particles are no longer selected by the second slit but by the first one and the density profile continues to raise linearly but with a different slope until $x_{B'}$:

\beq
pd(x)_{x \in[x_{C''} x_{I}]} & = & \frac{1}{L_2} \left[ \left(x - (X_s + \frac{X_s - \sx }{L_{1b}}(L_2 - L_{1b})) \right) \frac{L_{1b}}{L_2 - L_{1b}} \right]  \label{eq:thickPPprofile1}  \\
pd(x)_{x \in[x_{I} x_{B'}]} & = & \frac{1}{L_2} \left[ \left(x - (X_s + \frac{X_s - \sx }{L_{1a}}(L_2 - L_{1a})) \right) \frac{L_{1a}}{L_2 - L_{1a}} \right] \label{eq:thickPPprofile2}
\eeq 

Between $x_{B'}$ and $x_{D'}$ the particle density per unit length is constant (this interval is non-zero only when $x_{B'}  < x_{D'}$). 
\beq
pd(x)_{x \in[x_{B'} x_{D'}]} & = &\frac{2 \sx}{L_2} \label{eq:thickPPprofile3}
\eeq

The density then decreases linearly in two steps:

\beq
pd(x)_{x \in[x_{D'} x_{J}]} & = & \frac{1}{L_2} \left[ \left((X_s + \slitw \frac{X_s + \sx }{L_{1a}}(L_2 - {L_1a})) -x \right) \frac{L_{1a}}{ L_2 - L_{1a}} \right] \label{eq:thickPPprofile4} \\
pd(x)_{x \in[x_{J} x_{E''}]} & = & \frac{1}{L_2} \left[ \left((X_s + \slitw \frac{X_s + \sx }{L_{1b}}(L_2 - {L_1b})) -x \right) \frac{L_{1b}}{  L_2 - L_{1b}} \right]  \label{eq:thickPPprofile5}
\eeq

One can see that if the distance between the pepper-pots ($L_{1a}-L_{1b}$) tends toward 0 (i.e. $L_{1a} \rightarrow L_1$ and $L_{1b} \rightarrow L_1$) we find back the results derived in the case of a single thin pepper-pot (equations~\ref{eq:thinPPprofile1} to \ref{eq:thinPPprofile3}) (see also~\ref{sec:deltaPD}).

\begin{figure}[htbp]
\begin{center}
\begin{tabular}{cc}
\hspace{-1.4cm}\includegraphics[height=3cm]{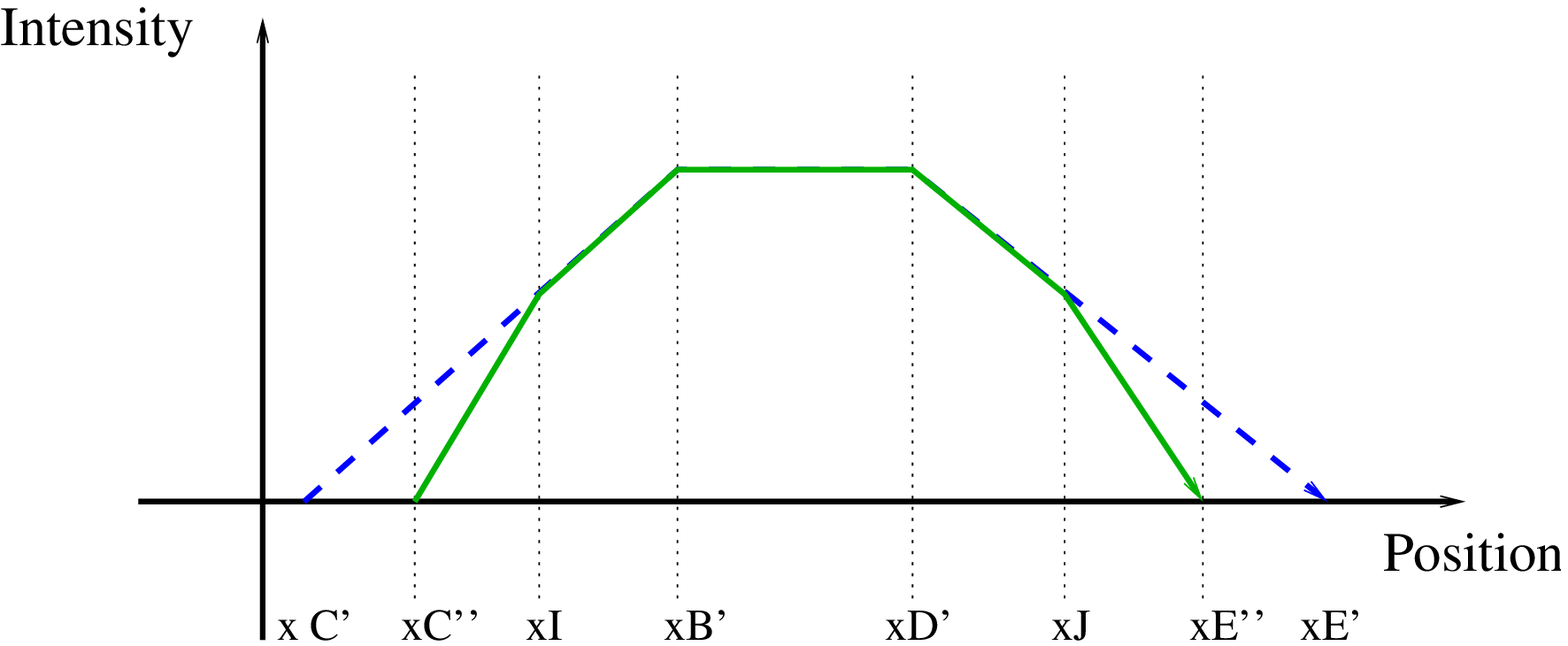}  &
\includegraphics[height=3cm]{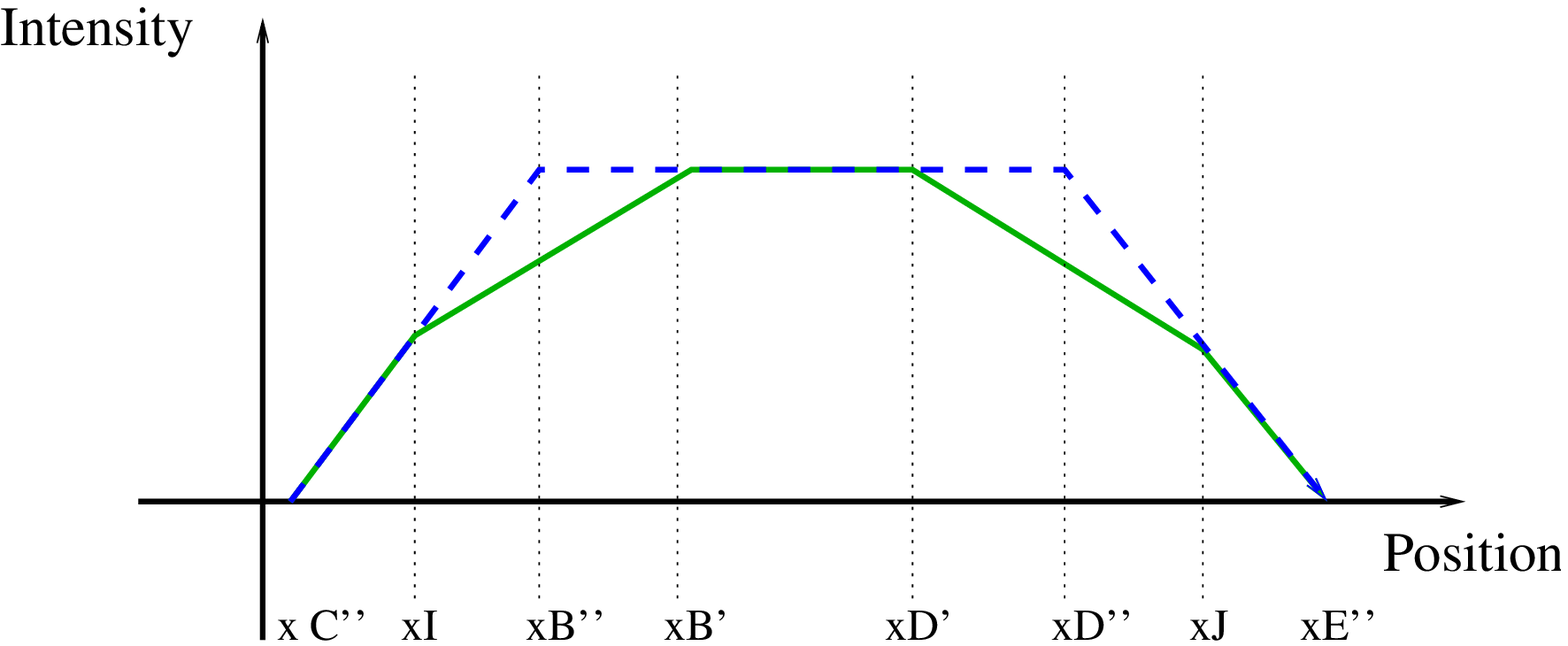}   \\
Thin PP located at entrance &
Thin PP located at exit \\
\end{tabular}
\caption{
Intensity profile expected on the screen for a beamlet after sampling by a long pepper-pot (in green plain line - equations \ref{eq:thickPPprofile1} to \ref{eq:thickPPprofile5}) and for an infinitely thin pepper-pot (in blue dashed line - equations~\ref{eq:thinPPprofile1} to \ref{eq:thinPPprofile3}) located either at the position of the front of the long pepper-pot (left figure) or at the position of the back (right figure).
\label{fig:beamProfile} }
\end{center}
\end{figure}

This profile is drawn on figure~\ref{fig:beamProfile}.

\subsection{One long pepper-pot}

The cases discussed in the previous two subsections can be extended to a continuous pepper-pot with an entry at $L_{1a}$  and an exit at $L_{1b}$.

The entry and exit of this pepper-pot are the two most extreme parts of the phase space that can be accepted by this pepper-pot. Hence all particles within the acceptance of both the entry and the exit of this pepper-pot will be within the total acceptance of this pepper-pot.


If the longitudinal dimension of the long pepper-pot ($\Delta L_{ba}$) is large enough very few particles will traverse the full thickness of the pepper-pots.  Some particles that are within the area swept by the pepper-pot but not in its full acceptance area may not be stopped and create a small background. However most of these particles will be scattered at large angle and thus not be observed by a screen located at a reasonable distance after the pepper-pot. Further considerations on the position of the pepper-pot and the screen are discussed in the next section.

Provided that the error discussed at equation~\ref{eq:areaRatioSmpl} is kept small, the resolution of the long pepper-pot will be similar to that of a thin pepper-pot as derived in equation~\ref{pepperpotRes}.

\section{Pepper-pot and screen separation in a drift space}
\label{sec:parametres}

We have shown that not all locations in the drift space are suitable for the installation of a long pepper-pot. This raises the question of where the pepper-pot and the screen used to observe the beamlets should be positioned.

\begin{figure}[htbp]
\begin{center}
\includegraphics[height=8cm]{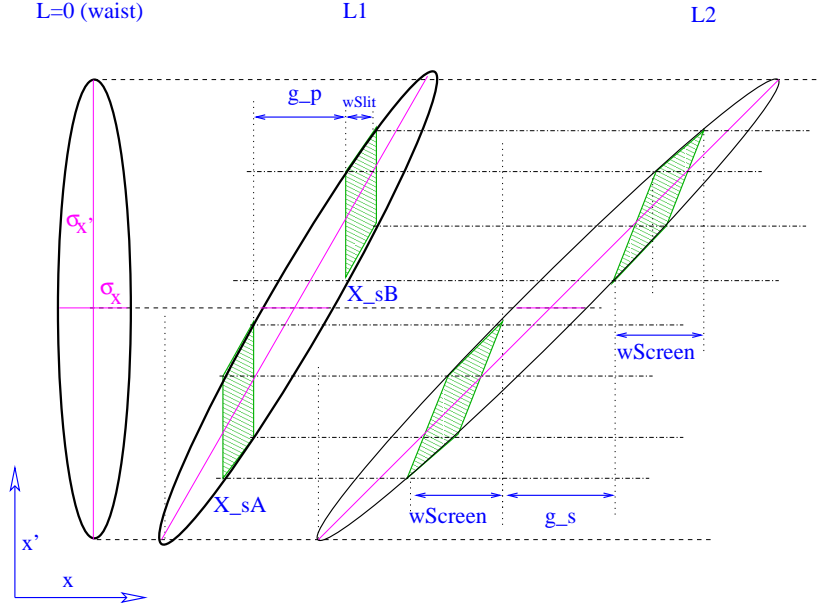} 
\caption{
Propagation in a drift space of a beam sampled by two slits of a pepper-pot.
\label{fig:slits} }
\end{center}
\end{figure}

Figure~\ref{fig:slits} shows how a beam sampled by two slits evolves in a drift space. Below we use the notation of that figure.

For the pepper-pot to be usable it is important that the beamlets do not overlap, hence the gap between the two beamlets must still exist at the position of the screen. This sets a condition on the size of the gap ($g_p$) between two slits of the pepper-pot which depends on the size of the gap at the screen $g_s$.

The size of this gap corresponds to the distance on the screen between the most diverging particles passing through the two slits.

Let $X_{sA}$ and $X_{sB}$ be the position of the two slits, with $X_{sA} < X_{sB}$. Using equation~\ref{eq:divergenceAtSourceWide} we can calculate the divergence ($x'_1A$ and $x'_1B$) of the particles passing each slit.

\beq
\frac{X_{sA} - \sx}{L_1} & < X_{1A}' < & \frac{X_{sA} + \sx + \slitw}{L_1} \nonumber \\
\frac{X_{sB} - \sx}{L_1} & < X_{1B}' < & \frac{X_{sB} + \sx + \slitw}{L_1} \nonumber 
\eeq

From this we can deduce the position ($x_{2A}$ and $x_{2B}$)of the particles on the second screen:

\beq
X_{sA} + \frac{X_{sA} - \sx}{L_1} (L_2 - L_1)& < x_{2A} < & X_{sA} + \slitw + \frac{X_{sA} + \sx + \slitw}{L_1} (L_2 - L_1)  \nonumber \\
X_{sB} + \frac{X_{sB} - \sx}{L_1} (L_2 - L_1)& < x_{2B} < & X_{sB} + \slitw + \frac{X_{sB} + \sx + \slitw}{L_1} (L_2 - L_1)  \nonumber
\eeq

Hence the size of the gap (with $X_{sB} - X_{sA} - \slitw = g_p $):
\beq 
g_s & = & X_{sB} + \frac{X_{sB} - \sx}{L_1} (L_2 - L_1) - \left[  X_{sA} + \slitw + \frac{X_{sA} + \sx + \slitw}{L_1} (L_2 - L_1)  \right]  \nonumber \\
& = & g_p \frac{L_2}{L_1} - 2 \sx \left( \frac{L_2}{L_1} - 1 \right)  \label{eq:gs}
\eeq

Hence to keep the gap open we must have:

\beq
g_s & > & 0 \nonumber \\
g_p & > &  2 \sx \left( 1 - \frac{L_1}{L_2} \right)  \label{eq:gapCond} \\
1 - \frac{g_p}{2 \sx} & < &  \frac{L_1}{L_2}  \label{eq:gapCondL}
\eeq

Unsurprisingly this equation shows that the further away from the waist the pepper-pot is, the bigger the gaps between the slits will have to be. It also shows that for beams with a large divergence (when $L \sxp>>\sx$) the size of the gaps depends only on the width of the beam at the waist.

Using equation~\ref{pepperpotRes} it is also possible to estimate the contribution of the slit width ($\mathcal{C}_{sw}$) to the pattern observed on the screen as a function of the position of the pepper-pot and of the screen:
\beq
\mathcal{C}_{sw} & = & \frac{\slitw L_2}{2\sx (L_2 - L_1) + \slitw L_2}  \nonumber \\
 & = & \frac{\slitw }{2\sx (1 - \frac{L_1}{L_2}) + \slitw} \label{eq:csw} \\ 
\mathcal{C}_{sw} & > & \frac{\slitw }{g_p + \slitw}
\eeq

Ideally one wants $C_{sw}$ as small as possible and this equation shows that to reach the best possible sensitivity to the beam size at the waist the screen should be positioned far away from the pepper-pot. However such positioning will also affect the the gap size required by the condition expressed at equation~\ref{eq:gapCond} and the resolution is ultimately decided by the relative size of the gaps and the slits width.

\begin{figure}[htbp]
\begin{center}
\includegraphics[height=6cm]{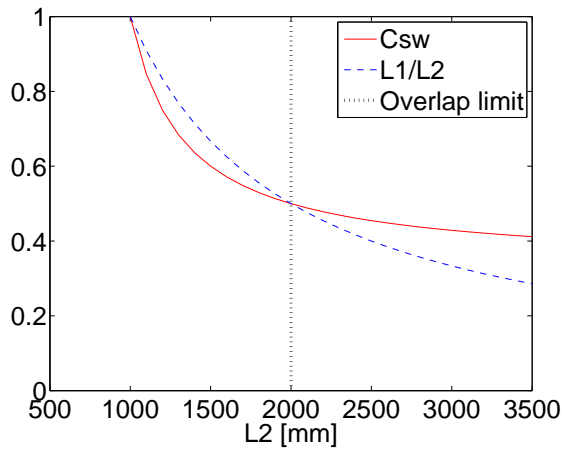}  
\end{center}
\caption{Value of $C_{sw}$ (equation~\ref{eq:csw}; red line) and $\frac{L1}{L2}$ (equation~\ref{eq:gapCondL}; blue dashed line) as a function of $L_2$ for $\slitw = 50 \um $; $\sx = 50\um $ ; $L_{1} = 1m $. The black dotted line shows the limit where $1 - \frac{g_p}{2 \sx} =  \frac{L_1}{L_2}$  as set by equation \ref{eq:gapCondL}.
\label{fig:csw} }
\end{figure}

\section{Numerical examples}

\begin{figure}[htbp]
\begin{center}
\begin{tabular}{cc}
\includegraphics[height=5cm]{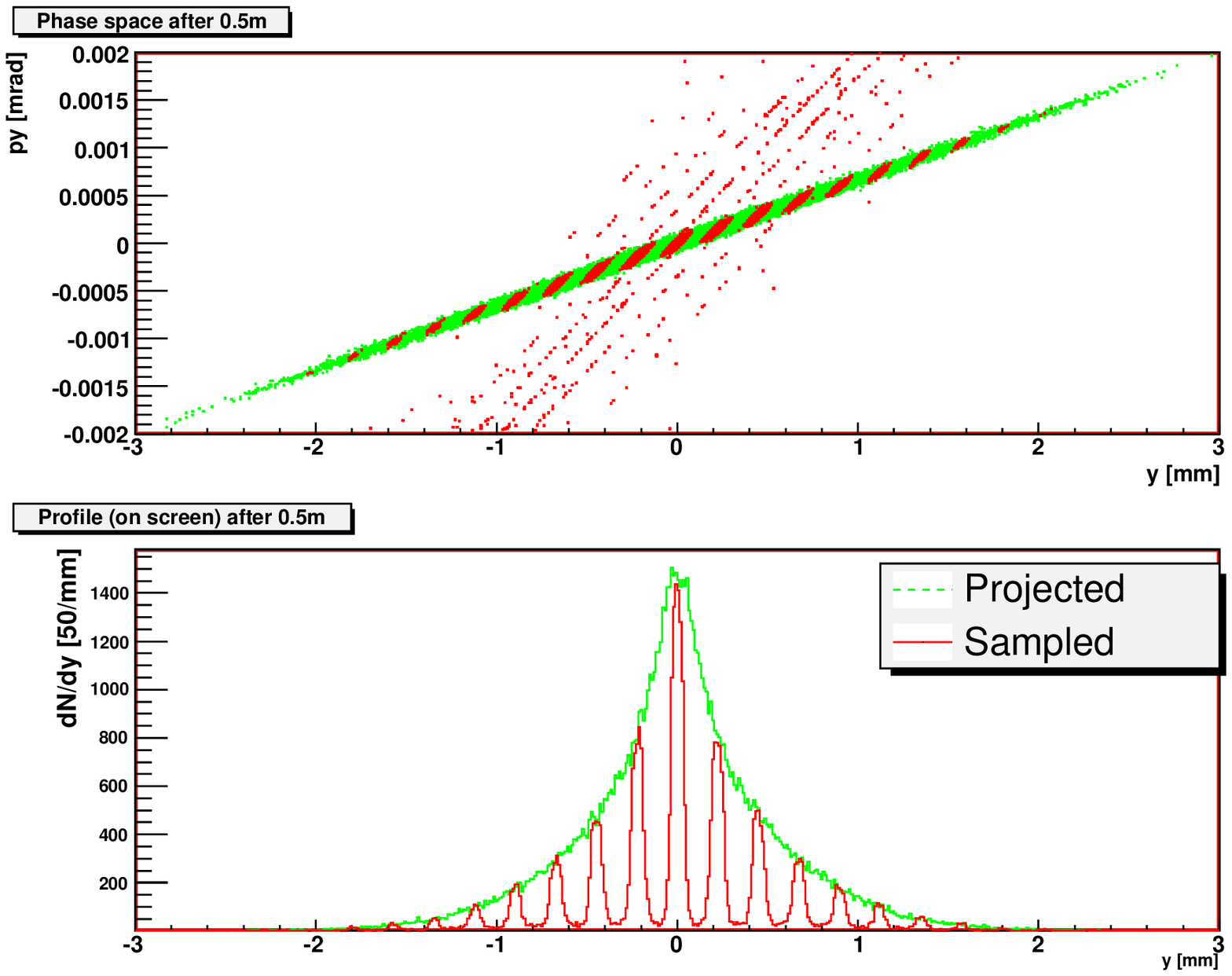}  &
\includegraphics[height=5cm]{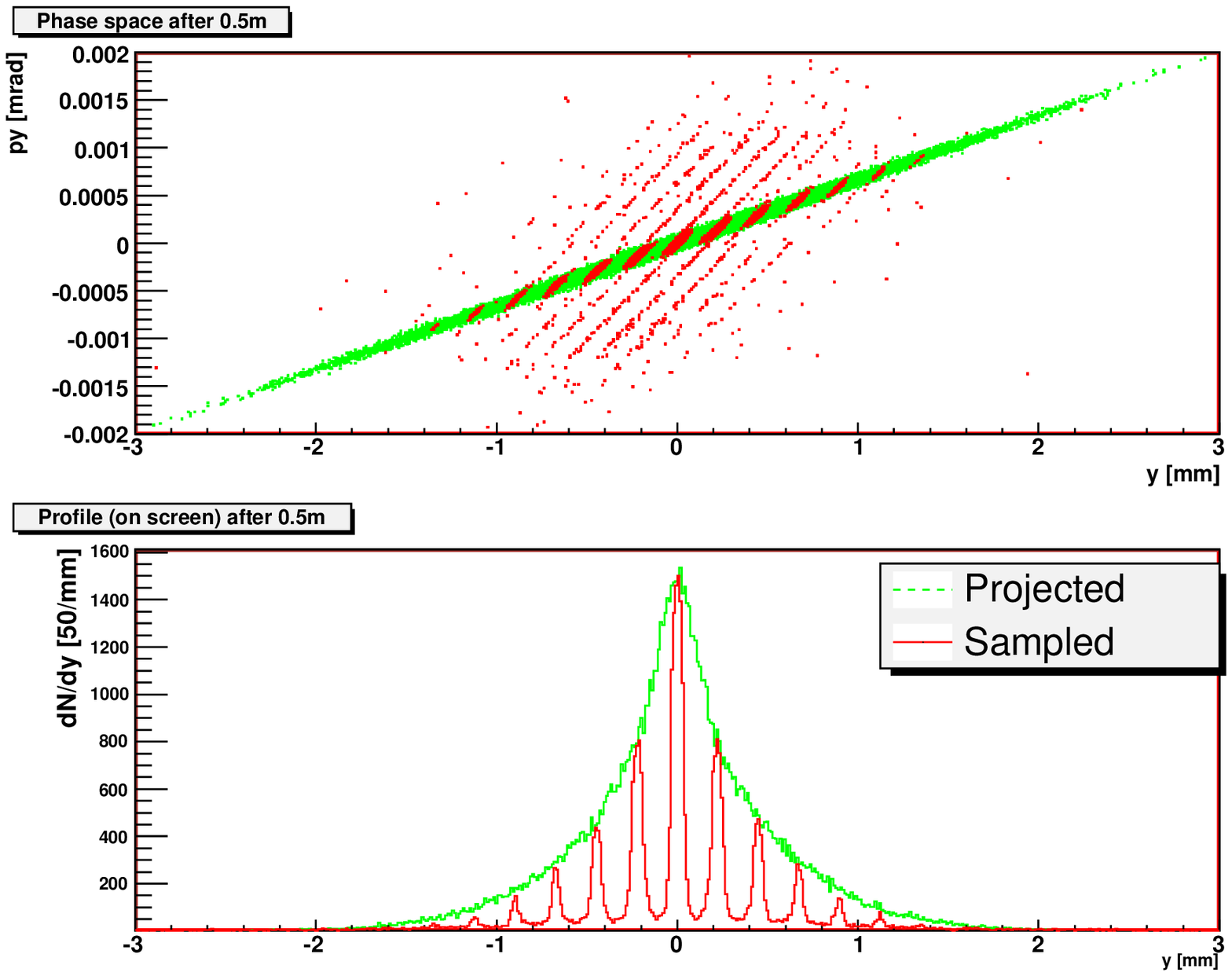}  \\
$ \Delta L_{ba} = 10\mm$; $\frac{\mathcal{A}_{s} - \mathcal{A}_{o}}{\mathcal{A}_{o}} = 4.5\%$ &  $ \Delta L_{ba} = 50\mm$; $\frac{\mathcal{A}_{s} - \mathcal{A}_{o}}{\mathcal{A}_{o}} = 22.5\%$ \\
\hspace*{2cm} & \\
\includegraphics[height=5cm]{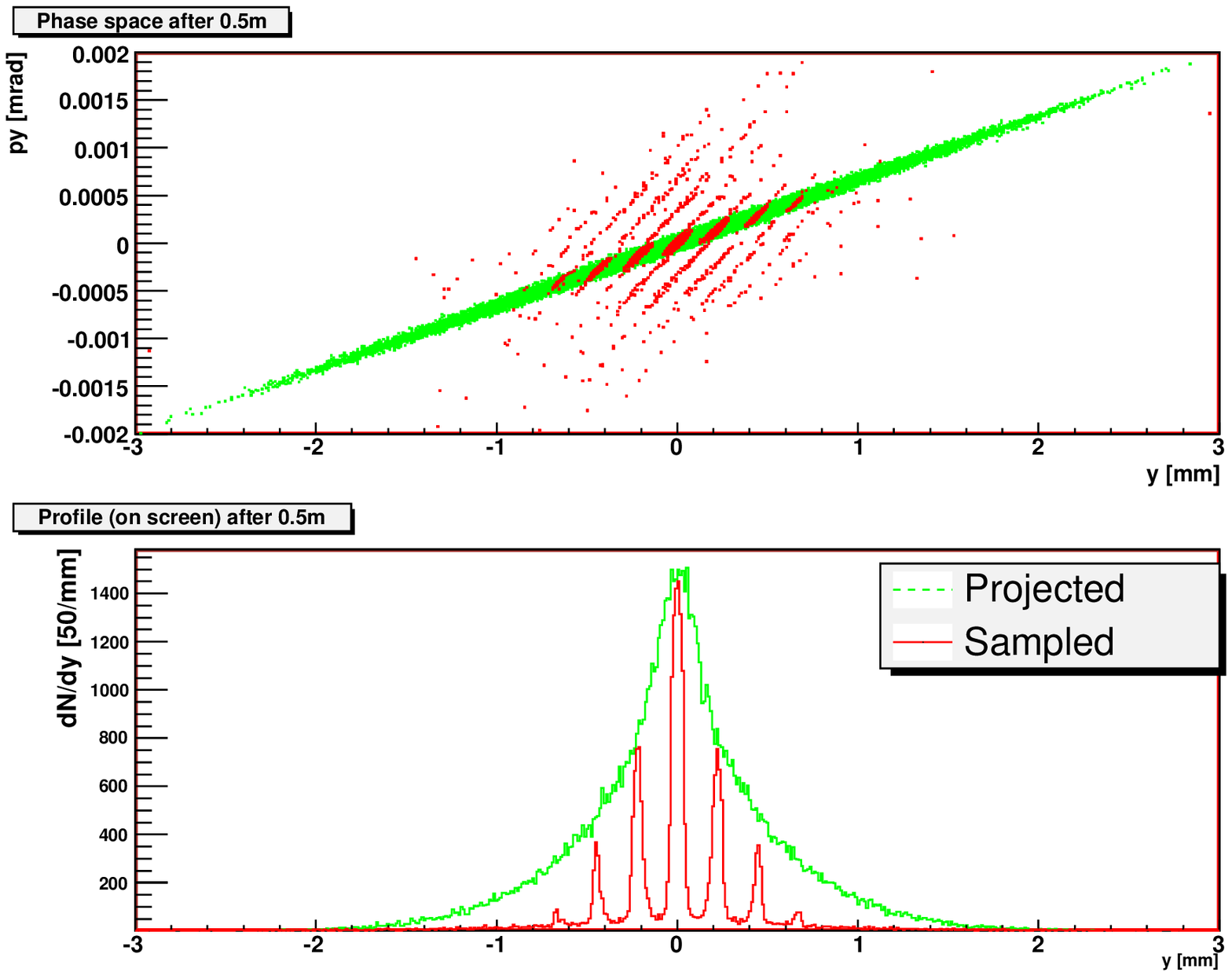}  &
\includegraphics[height=5cm]{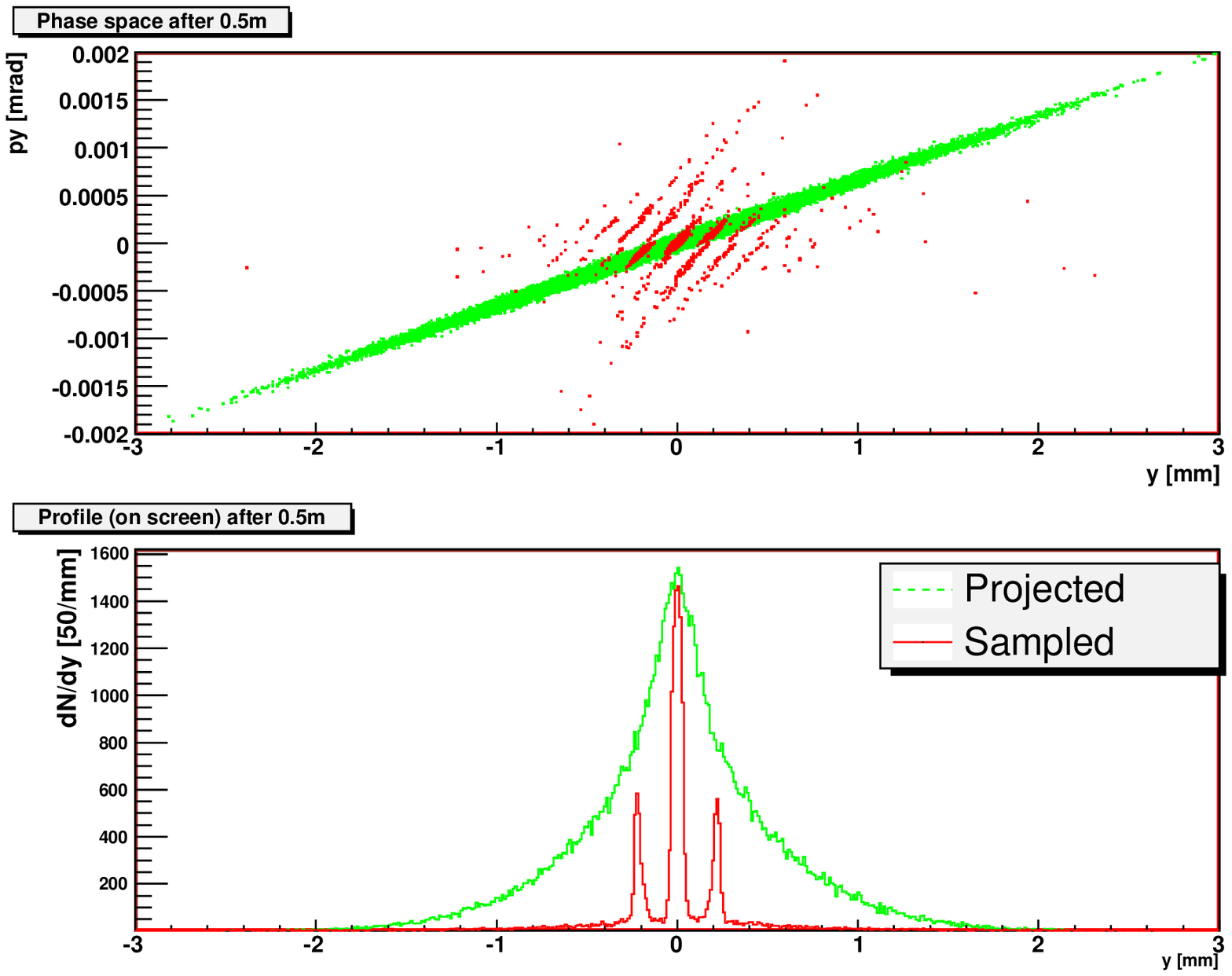}  \\
$ \Delta L_{ba} = 100\mm$; $\frac{\mathcal{A}_{s} - \mathcal{A}_{o}}{\mathcal{A}_{o}} = 45\%$ &  $ \Delta L_{ba} = 200\mm$; $\frac{\mathcal{A}_{s} - \mathcal{A}_{o}}{\mathcal{A}_{o}} = 90\%$ \\
gv acc\hspace*{2cm} & \\
\multicolumn{2}{c}{$\slitw = 50 \mu m $; $\sx = 50\um $ ;$\sxp = 0.5mrad $;}\\
\multicolumn{2}{c}{ $L_{1a} = 1m $; $L_2 = 1.5m$; 1\,GeV; 100~000 electrons}
\end{tabular}
\caption{
Reconstruction of the transverse phase space of a beam sampled by a long pepper-pot. In each pair of plots the top plot shows in green the original phase space of the electrons and in red the phase space of the electrons after the pepper-pot. The area of each red dot is proportionnal to the logarithm of the number of electrons in that particular part of the phase space.
In each pair the bottom plot shows the projection along the spatial component of the transverse phase space. In all these plots the beam energy is 1\GeV. The slit width is 50\um, the beam size at the waist is 100\um\ with a divergence of 1\mrad\ and the pepper-pot are located 1\m\ away from the beam waist. The screen is located 500\mm\ after the pepper-pot. For each plot 100~000 electrons were simulated with GEANT4.
The top left pair of plots corresponds to a pepper-pot length of 10\mm, the top right pair of plots corresponds to a pepper-pot length of 50\mm, the bottom left pair of plots corresponds to a pepper-pot length of 100\mm\ and the bottom right pair of plots correspond to a pepper-pot length of 200\mm.  For each plot the value of $\frac{\mathcal{A}_{s} - \mathcal{A}_{o}}{\mathcal{A}_{o}}$ as defined in equation~\ref{eq:areaRatioSmpl} is given. High energy photons (X-rays) are not shown on this figure.
\label{fig:GeantDepth} }
\end{center}
\end{figure}

GEANT4 has been used to validate these calculations by simulating the propagation in a long pepper-pot of 100~000~electrons with an energy of 1~GeV. The result of such simulation for various pepper-pot lengths is shown in figure~\ref{fig:GeantDepth}. It can be seen that for short lengths the phase space is not significantly clipped but for longer pepper-pot only the central beamlets remain. 

\begin{figure}[htbp]
\begin{center}
\begin{tabular}{cc}
\includegraphics[height=5cm]{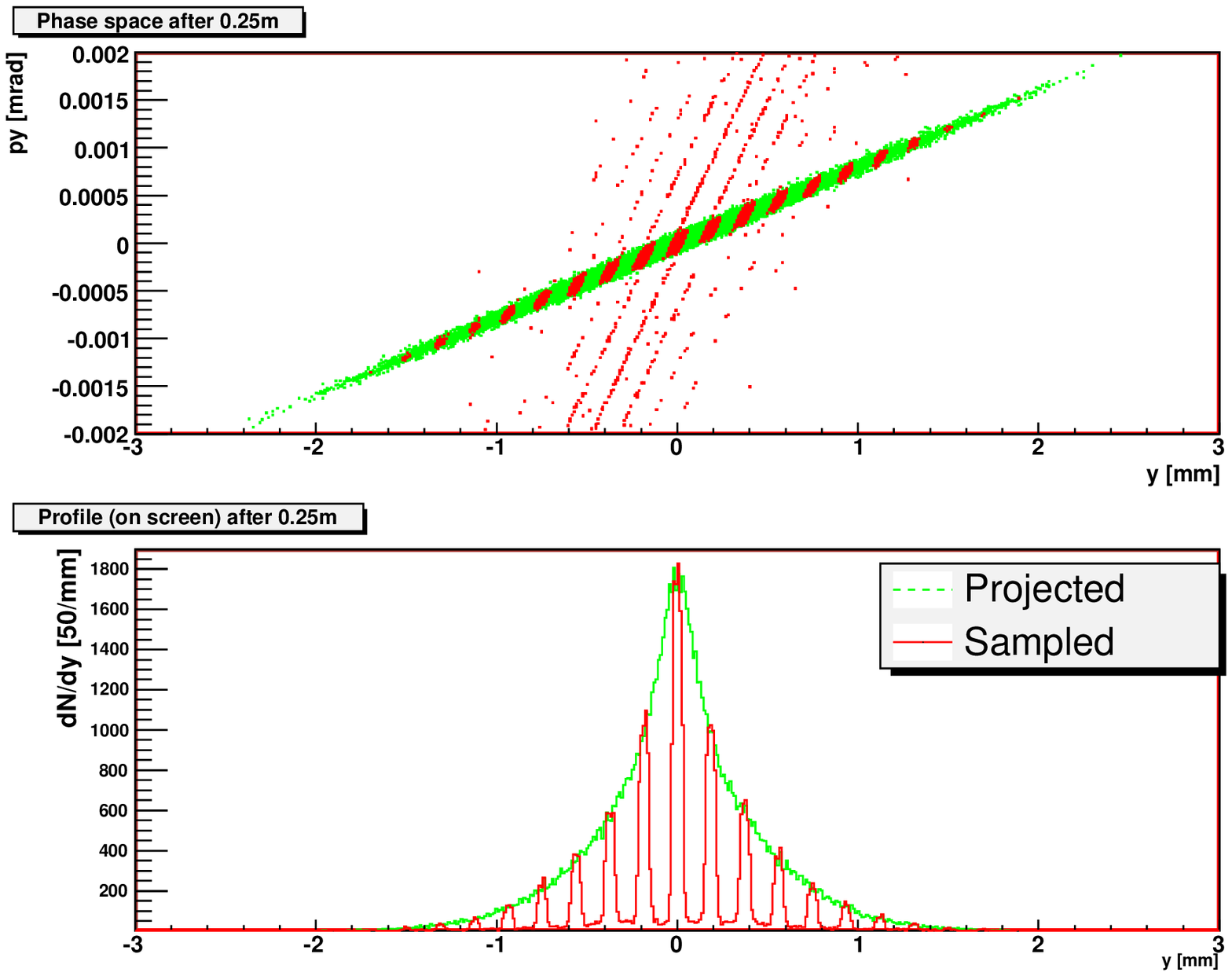}  &
\includegraphics[height=5cm]{acceptance_Air_Ta_sx50um_sxp05mrad_l10_ws50um_gp100um_xpm1_1E5_E1000_05m.eps}  \\
$L_2 - L_{1a} = 0.25\m$ ; $\frac{L_1}{L_2} = 0.8$ ; $\mathcal{C}_{sw} = 0.56$ & 
$L_2 - L_{1a} = 0.5\m$ ; $\frac{L_1}{L_2} = 0.67$ ; $\mathcal{C}_{sw} = 0.42$ \\
\includegraphics[height=5cm]{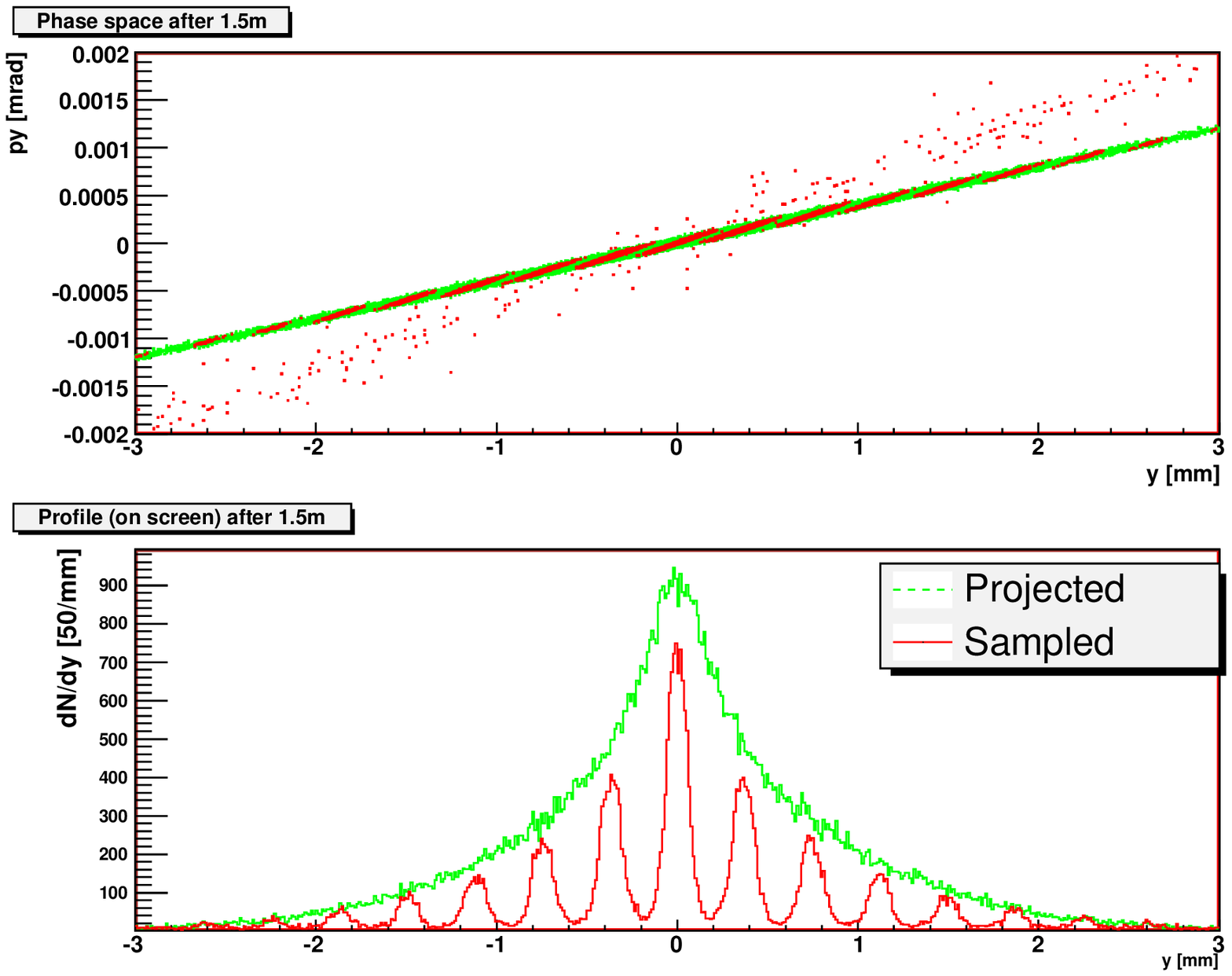}  &
\includegraphics[height=5cm]{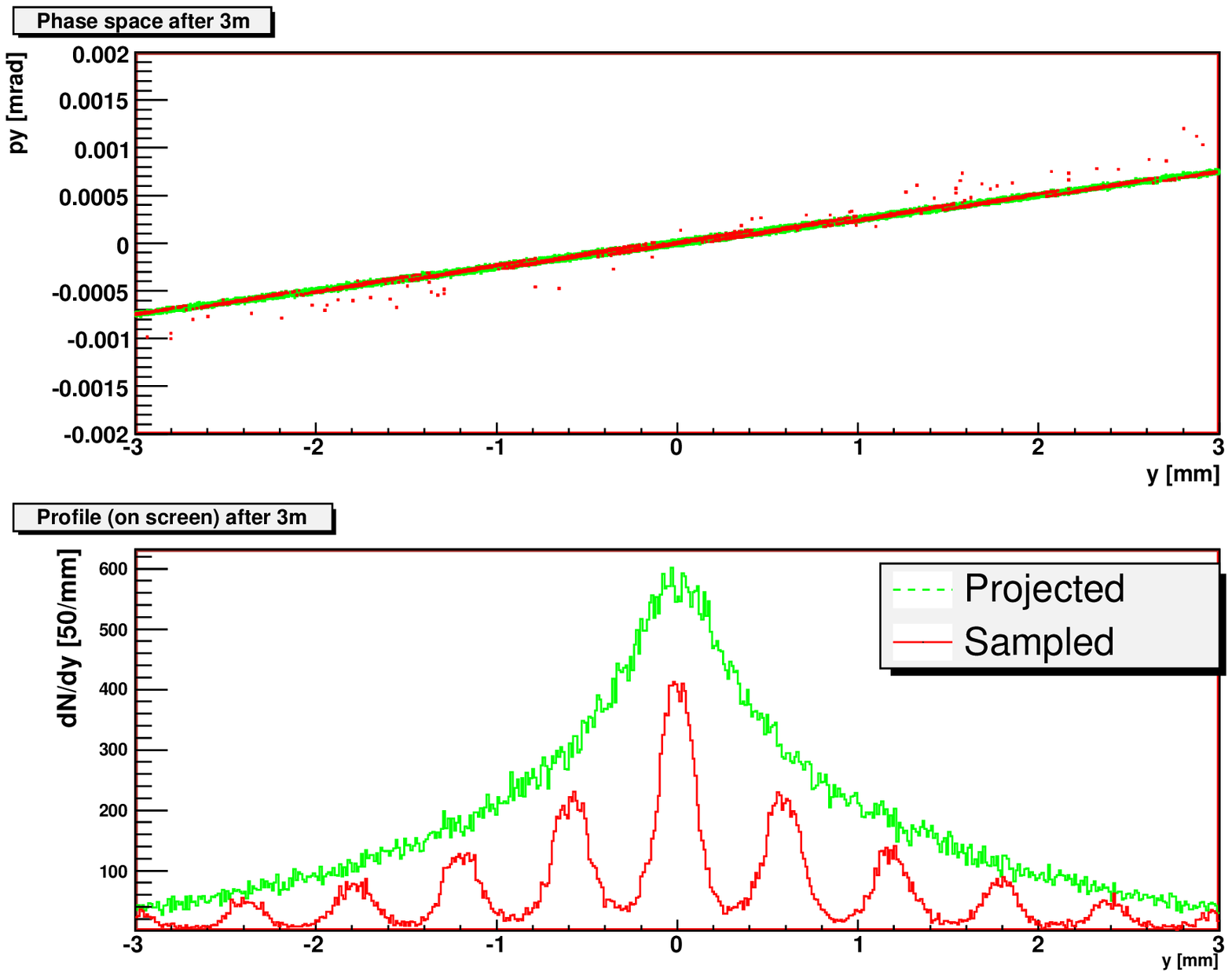}  \\
$L_2 - L_{1a} = 1.5\m$ ; $\frac{L_1}{L_2} = 0.4$ ; $\mathcal{C}_{sw} = 0.29$ & 
$L_2 - L_{1a} = 3\m$ ; $\frac{L_1}{L_2} = 0.25$ ; $\mathcal{C}_{sw} = 0.25$ \\

\hspace*{2cm} & \\

\multicolumn{2}{c}{$\Delta L_{ba} = 100\mm$;$\slitw = 50 \mu m $; $\sx = 100\um $ ; $L_{1a} = 1m $}\\
\multicolumn{2}{c}{$g_p = 100\um$;$ 1 - \frac{g_p}{2 \sx} = 0.5$; 1000\,MeV; 100~000 electrons}
\end{tabular}
\caption{
Reconstruction of the transverse phase space of a beam sampled by a long pepper-pot as a function of the distance of the imaging screen. In each pair of plots the top plot shows in green the original phase space of the electrons and in red the phase space of the electrons after the pepper-pot. The area of each red dot is proportionnal to the logarithm of the number of electrons in that particular part of the phase space. In each pair the bottom plot shows the projection along the spatial component of the transverse phase space. In all these plots the beam energy is 1\GeV. The slit width is 50\um, the gap size ($g_p$) is 100\um\  and the beam size at the waist is 100\um\   giving a ratio $ 1 - \frac{g_p}{2 \sx} = 0.5$ (see equation~\ref{eq:gapCondL})  
and the pepper-pot are located 1\m\ away from the beam waist. The length of the pepper-pot is 100\mm.  For each plot 100~000 electrons were simulated with GEANT4.
The top left pair of plots corresponds to an imaging screen located 250\mm\ after the pepper-pot, the top right pair of plots corresponds to an imaging screen located after 500\mm, the bottom left pair of plots corresponds to a screen located after 1.5\m\ and the bottom right pair of plots correspond to a screen located after 3\m. For each plot the value of $\frac{L_1}{L_2}$ as defined by equation \ref{eq:gapCondL} and of $\mathcal{C}_{sw}$ is given by equation~\ref{eq:csw} is given. High energy photons (X-rays) are not shown on this figure.
\label{fig:GeantPosition} }
\end{center}
\end{figure}

These simulations can also be used to show the range over which the screen can be positioned as shown in figure~\ref{fig:GeantPosition}. On this figure it is possible to see that when the conditions of equation~\ref{eq:gapCondL} are met, the beamlets do not overlap but when these conditions are not met anymore the beamlets do overlap.

\section{Conclusions}

Geometrical arguments have been used to demonstrate that a long pepper-pot measures the same phase space area as a thin pepper-pot provided that it is located far enough from the beam waist. We have given a formula that estimates the amount of phase space that is clipped by a long pepper-pot and shown numerical examples.
Similar arguments were used to calculate the effect of the position of the pepper-pot and of the imaging screen on the resolution of the system and two formulae were given, one to estimate the contribution of the slit width to the final measurement and the other one to estimate when the beamlets overlap.

In addition, we performed GEANT4 simulations to verify the results predicted by the analytical expressions. The overall results from this study demonstrate that a pepper-pot carefully designed should allow to measure in single shot the emittance of high energy electron beams in the MeV and GeV range such as those produced by laser-driven wakefield accelerators.

\section{Acknowledgements}

The author is grateful to the John Fell fund of the University of Oxford for their support with this work. The author is also grateful to George Doucas (University of Oxford) and Cyrille Thomas (Diamond Light Source) for their comments on the draft of this manuscript.

\appendix

\section{Derivation of equation~\ref{eq:areaRatio}}
\label{sec:deriveAreaRation}

Using the notation of figure~\ref{fig:parallelogram} we have:

\beq
\mathcal{A}_o & = & \mathcal{A}_{B'D''JE'C''I} \\
\mathcal{A}_s & = & \mathcal{A}_{B''D'JE''C'I}
\eeq

Now, let's consider the intersection $I$ of the left edge of the two areas. Because the particles at $I$ have passed both pepper-pots at $X_s$, the coordinates ($x_I$,$x'_I$) of $I$ have to satisfy the following:
\beq
x_{0I} + x'_I  L_{1a} & = & X_s = x_{0I} + x'_I L_{1b}
\eeq
where $x_{0I} = x_I - x'_I L_2$ hence:

\beq
x'_I=0 \\
x_I = X_s
\eeq

Then let's consider the coordinates of $B'$ at $L_2$. 
The particles at $B'$ were at the edge of the slit at $L_{1a}$ and, because they are now at the left edge of the ellipse, they were also at the left edge of the ellipse ($x=-\sx$) at the source. The divergence of these particles $x'_{B'}$ has not changed over the drift length, hence:

\beq
x'_{B'} & = & \frac{X_s + \sx}{L_{1a}} \\
x_{B'} & = & X_s +  \frac{X_s + \sx}{L_{1a}} (L_2  - L_{1a})
\eeq
(and similarly for $B''$).

The coordinates of the three vertices of the triangle $IB'B''$ are now known and hence it is possible to calculate its surface: $\mathcal{A}_{IB'B''}$

\beq
\mathcal{A}_{IB'B''} & = & \frac{1}{2} \left[ x'_{B''}  (x_{B'} - x_I) + x'_{B'}   (x_I -x_{B''} ) + x'_{I}  (x_{B''} -x_{I} ) \right] \nonumber \\
 & = & \frac{1}{2} \frac{(X_s + \sx)^2}{(L_{1a}L_{1b})} \Delta L_{ba} 
\eeq

Where $ \Delta L_{ba} =  L_{1b} - L_{1a}$.

We can find a similar relation for the other areas where the pepper-pots do not overlap. $\mathcal{A}_{IC'C''}$ can be found by replacing $-\sx$ with $+\sx$, $\mathcal{A}_{JD'D''}$ can be found by replacing  $X_s$ with $X_s + \slitw$ and $\mathcal{A}_{JE'E''}$ can be found by replacing $X_s$ with $X_s + \slitw$ and $-\sx$ with $+\sx$.

\beq
\mathcal{A}_{IC'C''} & = & \frac{1}{2} \frac{(X_s - \sx)^2}{L_{1a} L_{1b}}  \Delta L_{ba} \\
\mathcal{A}_{JD'D''} & = & \frac{1}{2} \frac{(X_s + \slitw + \sx)^2}{L_{1a} L_{1b}} \Delta L_{ba}  \\
\mathcal{A}_{JE'E''} & = & \frac{1}{2} \frac{(X_s + \slitw - \sx)^2}{L_{1a} L_{1b}} \Delta L_{ba} 
\eeq

From equations~\ref{eq:areaA} and~\ref{eq:areaB} we already know that:
\beq
\mathcal{A}_{B'D'E'C'} & = &  \frac{2\sx \slitw }{L_{1a}} \nonumber \\
\mathcal{A}_{B''D''E''C''} & = &  \frac{2\sx \slitw }{L_{1b}} \nonumber
\eeq

Hence:
\beq
\mathcal{A}_{o} & = & \mathcal{A}_{B'D'E'C'} - \mathcal{A}_{IC'C''} -  \mathcal{A}_{JD'D''}  \nonumber  \\
& = &  \frac{2\sx \slitw }{L_{1a}} - \frac{1}{2} \frac{(X_s - \sx)^2}{L_{1a} L_{1b}}  \Delta L_{ba} -   \frac{1}{2}  \frac{(X_s + \slitw + \sx)^2}{L_{1a} L_{1b}}  \Delta L_{ba}  \nonumber \\
& = & \frac{2\sx \slitw  L_{1b} -   \Delta L_{ba} \left[ (X_s - \sx)^2 + (X_s + \slitw + \sx)^2 \right]}{2 L_{1a} L_{1b}} \\
\mathcal{A}_{s} & = &  \mathcal{A}_{B'D'E'C'} + \mathcal{A}_{IB'B''} + \mathcal{A}_{JE'E''}  \nonumber  \\
& = & \frac{2\sx \slitw  L_{1b} +    \Delta L_{ba} \left[ (X_s + \sx)^2 + (X_s + \slitw - \sx)^2 \right]}{2 L_{1a} L_{1b}} \\
\frac{\mathcal{A}_{s} - \mathcal{A}_{o}}{\mathcal{A}_{o}} & = & \frac{ \Delta L_{ba} \left[ (X_s - \sx)^2 + (X_s + \slitw + \sx)^2 + (X_s - \sx)^2 + (X_s + \slitw + \sx )^2 \right] }{  2\sx \slitw L_{1b} +  \Delta L_{ba}  \left[ (X_s + \sx)^2 + (X_s + \slitw - \sx)^2 \right]  }   \nonumber 
\eeq

\section{Derivation of $\Delta(pd(x))$}
\label{sec:deltaPD}

It is possible to compute the difference $\Delta(pd(x))$ between the profile in the two cases (thin and thick pepper-pot) by subtracting equations~\ref{eq:thickPPprofile1} to~\ref{eq:thickPPprofile5} from equations~ \ref{eq:thinPPprofile1} to~ \ref{eq:thinPPprofile3}  (the notation used is that of figure~\ref{fig:parallelogram}).
Assuming that the thin pepper-pot is positioned at the same location than the entrance of the thick pepper-pot ($L_1 = L_{1a}$) we have:
\beq
\Delta(pd(x))_{x \in[x_{C'} x_{C''}]} &  =  &   \frac{1}{L_2} \left\{ \left[x - X_s -  (X_s - \sx )\frac{L_2 - L_{1a}}{L_{1a}} \right] \frac{L_{1a}}{L_2 - L_{1a}} \right\}  \\
\Delta(pd(x))_{x \in[x_{C''} x_{I}]} &  =  &  \frac{1}{L_2} \left\{ \left[ x - X_s - (X_s - \sx ) \left( \frac{L_2 - L_{1a}}{L_{1a}} - \frac{L_2 - L_{1b}}{L_{1b}} \right) \right] \right. \times \nonumber \\
& &  \left. \times \left(  \frac{L_{1a}}{L_2 - L_{1a}}  - \frac{L_{1b}}{L_2 - L_{1b}} \right) \right\} \\
\Delta(pd(x))_{x \in[x_{I} x_{B'}]} &  =  &  0 \\
\Delta(pd(x))_{x \in[x_{B'} x_{E'}]} &  =  &  0 \\
\Delta(pd(x))_{x \in[x_{E'} x_{J}]} &  =  &  0 \\
\Delta(pd(x))_{x \in[x_{J} x_{D''}]} &  =  &    \frac{1}{L_2} \left\{ \left[ \left(X_s + \slitw (X_s + \sx ) \left( \frac{L_2 - L_{1a}}{L_{1a}} - \frac{L_2 - L_{1b}}{L_{1b}} \right) \right) -x \right] \times \right. \nonumber \\ 
&  & \left. \times \left( \frac{L_{1a}}{L_2 - L_{1a}} -  \frac{L_{1a}}{L_2 - L_{1b}} \right] \right\}   \\
\Delta(pd(x))_{x \in[x_{D''} x_{D'}]} &  =  &   \frac{1}{L_2} \left\{ \left[ (X_s + \slitw (X_s + \sx ) \frac{L_2 - L_{1a}}{L_{1a}}) -x \right] \frac{L_{1a}}{L_2 - L_{1a}} \right\}   
\eeq

and assuming that the thin pepper-pot is positioned at the same location than the exit of the thick pepper-pot ($L_1 = L_{1b}$) we have:

\beq
\Delta(pd(x))_{x \in[x_{C''} x_{I}]} &  =  &  0 \\
\Delta(pd(x))_{x \in[x_{I} x_{B''}]} &  =  &  \frac{1}{L_2} \left\{ \left[ x - X_s - (X_s - \sx ) \left( \frac{L_2 - L_{1b}}{L_{1b}} - \frac{L_2 - L_{1a}}{L_{1a}} \right) \right] \right. \times \nonumber \\
& &  \left. \times \left(  \frac{L_{1b}}{L_2 - L_{1b}}  - \frac{L_{1a}}{L_2 - L_{1a}} \right) \right\} \\
\Delta(pd(x))_{x \in[x_{B''} x_{B'}]} &  =  &    \frac{1}{L_2} \left\{ (2 \sx) -  \right. \nonumber \\
& & \left. -  \left\{ \left[x - X_s -  (X_s - \sx )\frac{L_2 - L_{1a}}{L_{1a}} \right] \frac{L_{1a}}{L_2 - L_{1a}}  \right\}  \right\}  \\
\Delta(pd(x))_{x \in[x_{B'} x_{E'}]} &  =  &  0 \\
\Delta(pd(x))_{x \in[x_{E'} x_{E''}]} &  =  &   \frac{1}{L_2} \left\{ (2 \sx) \right. - \nonumber \\ 
& & \left. - \left[ \left((X_s + \slitw (X_s + \sx ) \frac{L_2 - L_{1a}}{L_{1a}}) -x \right) \frac{L_{1a}}{L_2 - L_{1a}} \right] \right\}\\ 
\Delta(pd(x))_{x \in[x_{E''} x_{J}]} &  =  &   \frac{1}{L_2} \left\{ \left[ \left(X_s + \slitw (X_s + \sx ) \left( \frac{L_2 - L_{1b}}{L_{1b}} - \frac{L_2 - L_{1a}}{L_{1a}} \right) \right) -x \right] \right. \times \nonumber \\ 
& & \times  \left. \left( \frac{L_{1b}}{L_2 - L_{1b}} -  \frac{L_{1a}}{L_2 - L_{1a}}  \right) \right\}   \\
\Delta(pd(x))_{x \in[x_{J} x_{D''}]} &  =  &  0 
\eeq

This gives the profiles shown on figure~\ref{fig:beamProfile}.

\bibliographystyle{elsarticle-num}
\bibliography{biblio}

\end{document}